\documentclass[twoside,twocolumn,9pt]{article}
\usepackage{extsizes}
\usepackage[super,sort&compress,comma]{natbib} 
\usepackage[version=3]{mhchem}
\usepackage[left=1.5cm, right=1.5cm, top=1.785cm, bottom=2.0cm]{geometry}
\usepackage{balance}
\usepackage{mathptmx}
\usepackage{sectsty}
\usepackage{graphicx} 
\usepackage{lastpage}
\usepackage[format=plain,justification=justified,singlelinecheck=false,font={stretch=1.125,small,sf},labelfont=bf,labelsep=space]{caption}
\usepackage{float}
\usepackage{fancyhdr}
\usepackage{fnpos}
\usepackage[english]{babel}
\addto{\captionsenglish}{%
  
}
\usepackage{array}
\usepackage{droidsans}
\usepackage{charter}
\usepackage[T1]{fontenc}
\usepackage[usenames,dvipsnames]{xcolor}
\usepackage{setspace}
\usepackage[compact]{titlesec}
\usepackage{hyperref}
\definecolor{cream}{RGB}{222,217,201}

\begin{document}

\pagestyle{fancy}
\thispagestyle{plain}
\fancypagestyle{plain}{
%%%HEADER%%%
\renewcommand{\headrulewidth}{0pt}
}
%%%END OF HEADER%%%

%%%PAGE SETUP - Please do not change any commands within this section%%%
\makeFNbottom
\makeatletter
\renewcommand\LARGE{\@setfontsize\LARGE{15pt}{17}}
\renewcommand\Large{\@setfontsize\Large{12pt}{14}}
\renewcommand\large{\@setfontsize\large{10pt}{12}}
\renewcommand\footnotesize{\@setfontsize\footnotesize{7pt}{10}}
\makeatother

\renewcommand{\thefootnote}{\fnsymbol{footnote}}
\renewcommand\footnoterule{\vspace*{1pt}% 
\color{cream}\hrule width 3.5in height 0.4pt \color{black}\vspace*{5pt}} 
\setcounter{secnumdepth}{5}

\makeatletter 
\renewcommand\@biblabel[1]{#1}            
\renewcommand\@makefntext[1]% 
{\noindent\makebox[0pt][r]{\@thefnmark\,}#1}
\makeatother 
\renewcommand{\figurename}{\small{Fig.}~}
\sectionfont{\sffamily\Large}
\subsectionfont{\normalsize}
\subsubsectionfont{\bf}
\setstretch{1.125} %In particular, please do not alter this line.
\setlength{\skip\footins}{0.8cm}
\setlength{\footnotesep}{0.25cm}
\setlength{\jot}{10pt}
\titlespacing*{\section}{0pt}{4pt}{4pt}
\titlespacing*{\subsection}{0pt}{15pt}{1pt}
%%%END OF PAGE SETUP%%%

%%%FOOTER%%%
\fancyfoot{}
\fancyfoot[LO,RE]{\vspace{-7.1pt}\includegraphics[height=9pt]{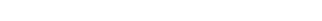}}
\fancyfoot[CO]{\vspace{-7.1pt}\hspace{11.9cm}\includegraphics{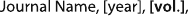}}
\fancyfoot[CE]{\vspace{-7.2pt}\hspace{-13.2cm}\includegraphics{head_foot/RF}}
\fancyfoot[RO]{\footnotesize{\sffamily{1--\pageref{LastPage} ~\textbar  \hspace{2pt}\thepage}}}
\fancyfoot[LE]{\footnotesize{\sffamily{\thepage~\textbar\hspace{4.65cm} 1--\pageref{LastPage}}}}
\fancyhead{}
\renewcommand{\headrulewidth}{0pt} 
\renewcommand{\footrulewidth}{0pt}
\setlength{\arrayrulewidth}{1pt}
\setlength{\columnsep}{6.5mm}
\setlength\bibsep{1pt}
%%%END OF FOOTER%%%

%%%FIGURE SETUP - please do not change any commands within this section%%%
\makeatletter 
\newlength{\figrulesep} 
\setlength{\figrulesep}{0.5\textfloatsep} 

\newcommand{\topfigrule}{\vspace*{-1pt}% 
\noindent{\color{cream}\rule[-\figrulesep]{\columnwidth}{1.5pt}} }

\newcommand{\botfigrule}{\vspace*{-2pt}% 
\noindent{\color{cream}\rule[\figrulesep]{\columnwidth}{1.5pt}} }

\newcommand{\dblfigrule}{\vspace*{-1pt}% 
\noindent{\color{cream}\rule[-\figrulesep]{\textwidth}{1.5pt}} }

\makeatother
%%%END OF FIGURE SETUP%%%

%%%TITLE, AUTHORS AND ABSTRACT%%%
\twocolumn[

%%% xxx 
\begin{@twocolumnfalse}
% {\includegraphics[height=30pt]{head_foot/PCCP}\hfill\raisebox{0pt}[0pt][0pt]{\includegraphics[height=55pt]{head_foot/RSC_LOGO_CMYK}}\\[1ex]
% \includegraphics[width=18.5cm]{head_foot/header_bar}}\par
% \vspace{1em}
% \sffamily
%%% xxx -commented out by KB 2025-01-31

\begin{tabular}{m{4.5cm} p{13.5cm} }

\includegraphics{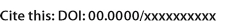} & \noindent\LARGE{\textbf{
Probing the structure of cyclic hydrocarbon molecules with X-ray-induced Coulomb explosion imaging
$^\dag$}} \\%Article title goes here ...
\vspace{0.3cm} & \vspace{0.3cm} \\

 & \noindent\large{
Kurtis D.~Borne\textit{$^{a,b}$}, 
Rebecca Boll\textit{$^{c}$}, 
Thomas M.~Baumann\textit{$^{c}$}, 
Surjendu Bhattacharyya\textit{$^{a,b}$},
Martin Centurion\textit{$^{d}$},  
Keyu~Chen\textit{$^{a}$}, 
Benjamin Erk\textit{$^{e}$},
Alberto De Fanis\textit{$^{c}$},
Ruaridh Forbes\textit{$^{b,f}$},
Markus Ilchen\textit{$^{c,e,g}$},
Edwin Kukk\textit{$^{h}$},
Huynh V.~S.~Lam\textit{$^{a}$},
Xiang Li\textit{$^{b}$}, 
Lingyu Ma\textit{$^{i}$},
Tommaso~Mazza\textit{$^{c}$},
Michael Meyer\textit{$^{c}$},
Terence Mullins\textit{$^{c,e,l}$},
J.~Pedro F.~Nunes\textit{$^{d}$},
Asami Odate\textit{$^{i}$},
Shashank Pathak\textit{$^{a}$},
Daniel~Rivas\textit{$^{c}$},
Philipp Schmidt\textit{$^{c}$},
%Bjoern Senfftleben\textit{$^{c}$},
Florian~Trinter\textit{$^{j}$},
Sergey~Usenko\textit{$^{c}$},
Anbu S.~Venkatachalam\textit{$^{a}$},
Enliang Wang\textit{$^{a,k}$},
Peter M.~Weber\textit{$^{i}$}, 
Till~Jahnke\textit{$^{c}$},
Artem~Rudenko\textit{$^{a}$}, 
and Daniel~Rolles\textit{$^{a,*}$}}
 \\%Author names go here ...

\includegraphics{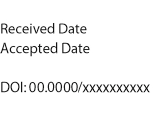} & \noindent\normalsize{

Coulomb explosion imaging (CEI) is a powerful experimental technique that maps a molecule's geometric structure onto the momenta of ionic molecular fragments produced by rapid multiple ionization. Here, we apply CEI induced by pulses from an X-ray free-electron laser in order to image and distinguish complex hydrocarbon isomers with the chemical formula \ce{C7H8}: toluene, cycloheptatriene, and 1,6-heptadiyne. The measured fragment-ion momentum distributions show discernible differences between the three isomers and provide signatures of specific carbon and hydrogen sites in the molecule. In contrast to previous work, we demonstrate that distinct 'marker atoms' are not strictly required for constructing a meaningful molecular frame of reference for the interpretation of the momentum-space data. Our work paves the way for tracking the ultrafast motion of nuclei during isomerization reactions in pure hydrocarbons.  
}

\end{tabular}

\end{@twocolumnfalse} 
 \vspace{0.6cm}
]
%%%END OF TITLE, AUTHORS AND ABSTRACT%%%

%%%FONT SETUP - please do not change any commands within this section
\renewcommand*\rmdefault{bch}\normalfont\upshape
\rmfamily
\section*{}
\vspace{-1cm}

%%%FOOTNOTES%%%

\footnotetext{\textit{$^{a}$~J.\ R.~Macdonald Laboratory, Department of Physics, Kansas State University, Manhattan, KS, USA}}
\footnotetext{\textit{$^{b}$~Linac Coherent Light Source, SLAC National Accelerator Laboratory, Menlo Park, CA, USA}}
\footnotetext{\textit{$^{c}$~European XFEL, Schenefeld, Germany}} %22869 Schenefeld
\footnotetext{\textit{$^{d}$~Department of Physics and Astronomy, University of Nebraska–Lincoln, Lincoln, NE, USA}}
\footnotetext{\textit{$^{e}$~Deutsches Elektronen-Synchrotron DESY, Hamburg, Germany}}
\footnotetext{\textit{$^{f}$~Department of Chemistry, University of California, Davis, CA, USA}}
\footnotetext{\textit{$^{g}$~Department of Physics, University of Hamburg, Hamburg, Germany}}
\footnotetext{\textit{$^{h}$~Department of Physics and Astronomy, University of Turku, Turku, Finland}}
\footnotetext{\textit{$^{i}$~Department of Chemistry, Brown University, Providence, RI, USA}} %Rhode Island
\footnotetext{\textit{$^{j}$~Molecular Physics, Fritz-Haber-Institut der Max-Planck-Gesellschaft, Berlin, Germany}}
\footnotetext{\textit{$^{k}$~Hefei National Research Center for Physical Sciences at the Microscale and Department of Modern Physics, University of Science and Technology of China, Hefei, China}} %Hefei 230026,
\footnotetext{\textit{$^{l}$~The Hamburg Centre for Ultrafast Imaging, Universität Hamburg, Hamburg, Germany}}
\footnotetext{\textit{$^{*}$~email: rolles@ksu.edu}}

%Please use \dag to cite the ESI in the main text of the article.
%If you article does not have ESI please remove the the \dag symbol from the title and the footnotetext below.
\footnotetext{\dag~Supplementary Information available: [details of any supplementary information available should be included here]. See DOI: 10.1039/cXCP00000x/}

%\footnotetext{\ddag~Additional footnotes to the title and authors can be included \textit{e.g.}\ `Present address:' or `These authors contributed equally to this work' as above using the symbols: \ddag, \textsection, and \P. Please place the appropriate symbol next to the author's name and include a \texttt{\textbackslash footnotetext} entry in the correct place in the list.} 
     % commented out by KB 19-01-2024

%%%END OF FOOTNOTES%%%

%%%MAIN TEXT%%%%

\section{Introduction}

    Coulomb explosion imaging (CEI) is an experimental technique that encodes information about a molecule's nuclear geometry in the momenta of ionic fragments produced when the molecule breaks apart after being multiply ionized.
    The key to CEI is \emph{rapid} removal of multiple electrons, thus populating a highly charged cationic state, on which the molecule fragments -- ideally completely into individual atomic ions. \cite{vager_Coulomb_1989,pitzer_direct_2013,boll_x-ray_2022}
    If enough electrons are removed and the molecule does not undergo significant structural rearrangement during the ionization process, the measured asymptotic momenta of the fragment ions provide information about the molecular geometry prior to ionization. This information can be extracted, e.g., by comparing the measured momenta to predictions of a Coulomb explosion model, thus providing a link between the real-space molecular geometry and the observed fragment-ion momentum distribution. \cite{bhattacharyya_strong-field-induced_2022,li_coulomb_2022,lam_differentiating_2024,green_submitted,lam_SO2,zhou_coulomb_2020}
    Various methods can be employed for the ionization step, including colliding molecular ions with thin foils \cite{vager_Coulomb_1989,gemmell_determining_1980,herwig_metastable_2013}, ion-neutral cross-beam collisions\cite{neumann_fragmentation_2010,schmidt_spatial_2012,wang_ultrafast_2020, yuan_coulomb_2024}, and different ionizing radiation sources such as femtosecond lasers\cite{stapelfeldt_time-resolved_1998,schouder_laser-induced_2022,bhattacharyya_strong-field-induced_2022,lam_differentiating_2024, crane2023molecular,hishikawa_visualizing_2007,pitzer_direct_2013,endo_capturing_2020,mcdonnell_ultrafast_2020}, synchrotrons\cite{ pitzer_absolute_2016,ablikim_identification_2016}, or free-electron lasers \cite{boll_x-ray_2022,li_coulomb_2022, jahnke_direct_2025,  green_submitted,li2025imaging}. Similarly, the generated fragment ions can either be imaged in a non-coincident manner \cite{schouder_laser-induced_2022,christensen_dynamic_2014, amini_photodissociation_2018, pickering_communication_2016,burt2017coulomb,allum_coulomb_2018,cheng_multiparticle_2023} or by performing coincidence measurements in order to extract correlated information on a single-molecule level \cite{vager_Coulomb_1989,pitzer_direct_2013,herwig_metastable_2013,boll_x-ray_2022,bhattacharyya_strong-field-induced_2022,li_coulomb_2022, lam_differentiating_2024,green_submitted,lam_SO2,neumann_fragmentation_2010,schmidt_spatial_2012,wang_ultrafast_2020, yuan_coulomb_2024,endo_capturing_2020, mcdonnell_ultrafast_2020,pitzer_absolute_2016,ablikim_identification_2016,jahnke_direct_2025,li2025imaging,venkatachalam2025exploiting, Richard2025}.  

    Several studies used CEI specifically to differentiate molecular isomers.
    For example, CEI induced by soft X-ray synchrotron radiation was used to distinguish geometric isomers and conformers of halogenated hydrocarbons \cite{pitzer_absolute_2016, ablikim_identification_2016,pathak_differentiating_2020}.
    However, a single photoabsorption in the soft X-ray range generates only a few charges per molecule -- typically two to three, unless heavy atoms are present that enhance ionization. For polyatomic molecules in which the number of atoms exceeds the number of charges generated, single photoabsorption therefore does not lead to a complete breakup into individual atomic ions. In contrast, using intense soft X-ray pulses from an X-ray free-electron laser (XFEL) can result in very high charge states through multiple inner-shell ionizations and Auger decay cascades, thus triggering a complete breakup into atomic ions even for molecules with ten or more atoms.
    This has enabled complete CEI of CH$_3$I \cite{li_coulomb_2022} and of several heterocyclic molecules containing strong absorption sites such as iodine and sulfur. \cite{boll_x-ray_2022,jahnke_direct_2025,green_submitted} Complete Coulomb explosion can also be induced by intense near-infrared femtosecond laser pulses, which was used, e.g., to image the structure of molecules such as a biphenyl derivative \cite{hansen_torsion_2012,christensen_dynamic_2014}, bromochlorofluoromethane \cite{pitzer_direct_2013}, tribromomethane \cite{bhattacharyya_strong-field-induced_2022}, furan \cite{wang_time-resolved_nodate}, and several other organic molecules containing unique marker atoms \cite{lam_differentiating_2024,venkatachalam2025exploiting}.

    When used in a pump-probe scheme with femtosecond near-infrared or X-ray pulses, CEI can also be used to visualize structural changes in gas-phase molecules.\cite{lam_SO2,jahnke_direct_2025,wang_time-resolved_nodate,stapelfeldt_time-resolved_1998,crane2023molecular,endo_capturing_2020,mcdonnell_ultrafast_2020,li2025imaging,hansen_torsion_2012,christensen_dynamic_2014}
    Compared to other ultrafast gas-phase molecular imaging techniques such as ultrafast electron diffraction or hard X-ray scattering \cite{minitti_imaging_2015,wolf_photochemical_2019,ruddock_deep_2019}, CEI has a few notable differences that can be advantageous for certain types of studies. For example, obtaining structural information on light atoms such as hydrogen is challenging in electron diffraction and X-ray scattering due to their small atomic scattering form factors as compared to heavier atoms with higher nuclear charge \cite{hubbell_atomic_1975,centurion_ultrafast_2022,moreno_carrascosa_ab_2019}, making it extremely difficult to image the positions of hydrogen atoms in larger hydrocarbons.
    %may omit significant information of the nuclear dynamics in ring-reconfiguration reactions of hydrocarbons.
    %Additionally, the scattering signal only provides information on internal coordinate distances, while information about bond angles cannot be explicitly extracted \cite{hegazy_applying_2023}.
    % Thus, despite being extremely powerful, these ultrafast scattering methods provide only a limited picture of the molecular geometry., 
    Coulomb explosion imaging, on the other hand, has the same sensitivity to light and heavy species and is inherently sensitive to the full 3D structure, including all bond angles, not just atomic pair distances.
    Nevertheless, the direct extraction of position-space geometries from measured momentum-space data is far from trivial. So far, it has been performed routinely only for small molecules consisting of 2-3 atoms and represents a significant challenge for larger molecules.\cite{li_coulomb_2022,li_AI_submitted}  
    % provide a more intuitive and complete representation of molecular geometries and their ultrafast changes upon photoexcitation, and Coulomb explosion has been shown to be a promising candidate. 
    In addition, for larger molecules, CEI has been successfully applied in the past mainly to systems with unique 'marker' atoms. These markers facilitate the definition of a molecular frame of reference for the interpretation of the measured results \cite{jahnke_direct_2025,green_submitted,endo_capturing_2020}. An application of CEI to pure hydrocarbons with more than two carbon atoms has been little explored to date.

	In the present work, we study Coulomb explosion of three different isomers with the chemical formula \ce{C7H8}: toluene, cycloheptatriene, and 1,6-heptadiyne, shown in Fig.~\ref{fgr:newton_plot_main})A-C. 
    %They vary in cyclic and non-cyclic geometry, the number of carbons in a cyclic group, and the possibility of different conformer geometries. 
    Toluene consists of a methyl group attached to a phenyl ring; cycloheptatriene is a non-planar seven-membered ring; and 1,6-heptadieyene consists of a seven-membered carbon chain that has several conformers, as discussed in more detail in section \ref{sec:hep}. 
    % These molecules provide some key insights into CEI of biologically relevant pure hydrocarbons.
    These molecules have several interesting and challenging features with regards to CEI: First, there is no preferential absorption site within these molecules. Second, they lack unique ‘marker atoms' that could serve to define the molecular frame. 
    %As mentioned above, this aspect poses additional complication to the interpretation of the measured multi-dimensional fragment-ion momentum-space data, since such marker atoms are typically employed to define an intuitive molecular frame of references in the data analysis. 
    Third,  
    % These \ce{C7H8} isomers are all commercially available, easy to bring into the gas phase, and 
    their different geometries provide an opportunity to investigate if CEI can distinguish different isomers that could be formed, e.g., as reaction products of a ring-opening or ring-reconfiguration reaction of pure hydrocarbons. In the present work, we address the impact of these aspects for momentum imaging.

\section{Methods}

    \subsection{Experiment}
    
    The experiment was performed at the Small Quantum Systems (SQS) scientific instrument of the European XFEL\cite{decking_mhz-repetition-rate_2020} using the cold target recoil ion momentum spectroscopy (COLTRIMS) reaction microscope (REMI) endstation. This experimental setup has been described in previous publications \cite{kastirke_double_2020,kastirke_photoelectron_2020,jahnke_inner-shell-ionization-induced_2021,li_coulomb_2022,boll_x-ray_2022} and is only summarized here. 
    The liquid samples were contained in a stainless-steel sample reservoir at room temperature, and their vapor was delivered into the vacuum chamber as a molecular beam expanded through a 200 $\mu$m flat nozzle without the use of a carrier gas. 
    The molecular beam was collimated by passing it through four differential pumping stages, each separated by a molecular beam skimmer or aperture. 

    The ion spectrometer used in this experiment was 
    %configured to have a 15-cm extraction length with a 4~kV voltage drop for a uniform spectrometer field of 235 V/cm.
    similar to that used in Refs.~\cite{boll_x-ray_2022, Richard2025}. It consisted of an acceleration region with a length of 50~mm, followed by a drift region with a length of 130~mm with no mesh in between these two regions. The electric field at the target region was set to approximately 325~V/cm.
    The ions were detected on a 120-mm diameter MCP with an applied voltage of $-2.5$~kV across the chevron stack, followed by a hexagonal delay-line anode assembly held at ground.
    The ions originating from each X-ray pulse were measured in coincidence. The analog detector signals were recorded %at 10\,Hz with 
    using fast (1.8 GS/s) analog-to-digital converters.
    An average of 4 ion hits per X-ray pulse were recorded, of which an average of 2 ion hits per pulse originated from the molecular beam.
    
    The European XFEL operated with self-amplified spontaneous emission (SASE) at a base repetition rate of 10 Hz, providing bursts of electron bunches with an intra-train repetition rate of 1.1\,MHz.
    To acquire the data shown here, every sixth electron pulse was used to generate 1.5-keV X-ray pulses, yielding an effective rate of 600 soft X-ray pulses per second for the experiment.
    %, i.e., 60 pulses per 10-Hz train.
    %The photon energy was $\sim 1.5$ keV.
    The average single-shot pulse energy was up to 4 mJ, as measured by a gas monitor detector upstream of the beamline optics.\cite{xgmd} Given an estimated beamline transmission of 80\%,\cite{Mazza_2023} this resulted in an average pulse energy of up to 3.2~mJ  at the interaction point.
    The X-ray pulses were focused using a pair of Kirkpatrick-Baez mirrors, and the focus spot size was estimated to be 1.4 $\mu$m diameter FWHM.
    The estimated upper limit of the X-ray pulse duration based on the electron bunch charge of 250 pC in the accelerator is 25\,fs (FWHM). However, dedicated tuning of the undulator settings with feedback from the SASE3 grating spectrometer before the data taking suggests significantly shorter pulse durations based on spectral analysis for the present experiment.
% https://doi.org/10.22003/XFEL.EU-DATA-002448-00

    \subsection{Coulomb explosion simulations}
    \label{sec:sim}

    Numerical simulations of the Coulomb explosion process were performed by solving $3N$ coupled differential equations for classical point particles of mass $ m_{i}$ and charge $ q_{i}$, located at positions $\vec{r_{i}}$, undergoing mutual Coulomb repulsion. The simulations assumed an instantaneous charge-up to the final charge state, and the initial coordinates of the ions were defined by the molecular geometry as described below. 
    The system of equations was defined by Newton's second law under the Coulombic force:
    \begin{equation}
        m_{i}\frac{d^{2}}{dt^{2}}\vec{r_{i}}(t)-\sum_{j=1}^{N} q_{i}q_{j}\frac{\vec{r_{i}}(t)-\vec{r_{j}}(t)}{\left( \vec{r_{i}}(t)-\vec{r_{j}}(t) \right)^{3/2}}=0 
    \end{equation}

    To model the experimental results presented in the main text, all carbon atoms were doubly charged ($q=2$), and all hydrogen atoms were singly charged ($q=1$) in order to match the coincidence channel that was selected in the data analysis.
    To simulate the influence of the inherent uncertainty of the atomic positions due to zero-point quantum fluctuations~\cite{Richard2025} on the momentum imaging, several thousand geometries with initial positions reflecting the zero-point energy vibrational distributions ('Wigner distribution') were generated.
    First, the neutral electronic ground-state geometry of each molecule was determined using the universal force field molecular optimization method with the Babel extension in Avogadro-2 \cite{hanwell_avogadro_2012,oboyle_open_2011}.
    This geometry was then used in the Newton-X software\cite{barbatti_newton-x_2014} to sample 10,000 geometries of the distribution in the ground state at 0~K. While several earlier CEI studies have shown that sampling the Wigner distribution typically leads to a significantly narrower spread of the simulated ion momenta than what is observed experimentally\cite{bhattacharyya_strong-field-induced_2022,lam_differentiating_2024,green_submitted,venkatachalam2025exploiting}, the narrower features in the simulated momentum images make it easier to identify the contributions from different molecular reference frames, and we therefore chose not to add additional broadening to the simulations.

\section{Results and Discussion}
   
\subsection{Molecular-frame momentum-space results}

In this section, we first discuss the experimental results and their comparison for the three isomers in general terms, followed by a discussion of the simulation results for the three isomers. In sections \ref{sec:toluene} through \ref{sec:hep}, we revisit the results for each isomer in detail and discuss how the Coulomb explosion images can be improved by using additional angular correlations to better define a molecular reference frame. Finally, while the aforementioned sections focus on carbon ions, section \ref{sec:hydrogens} presents experimental and simulated results for the emitted H$^+$ ions.

    \begin{figure*}[t]
        \centering
        \includegraphics[height=13cm]{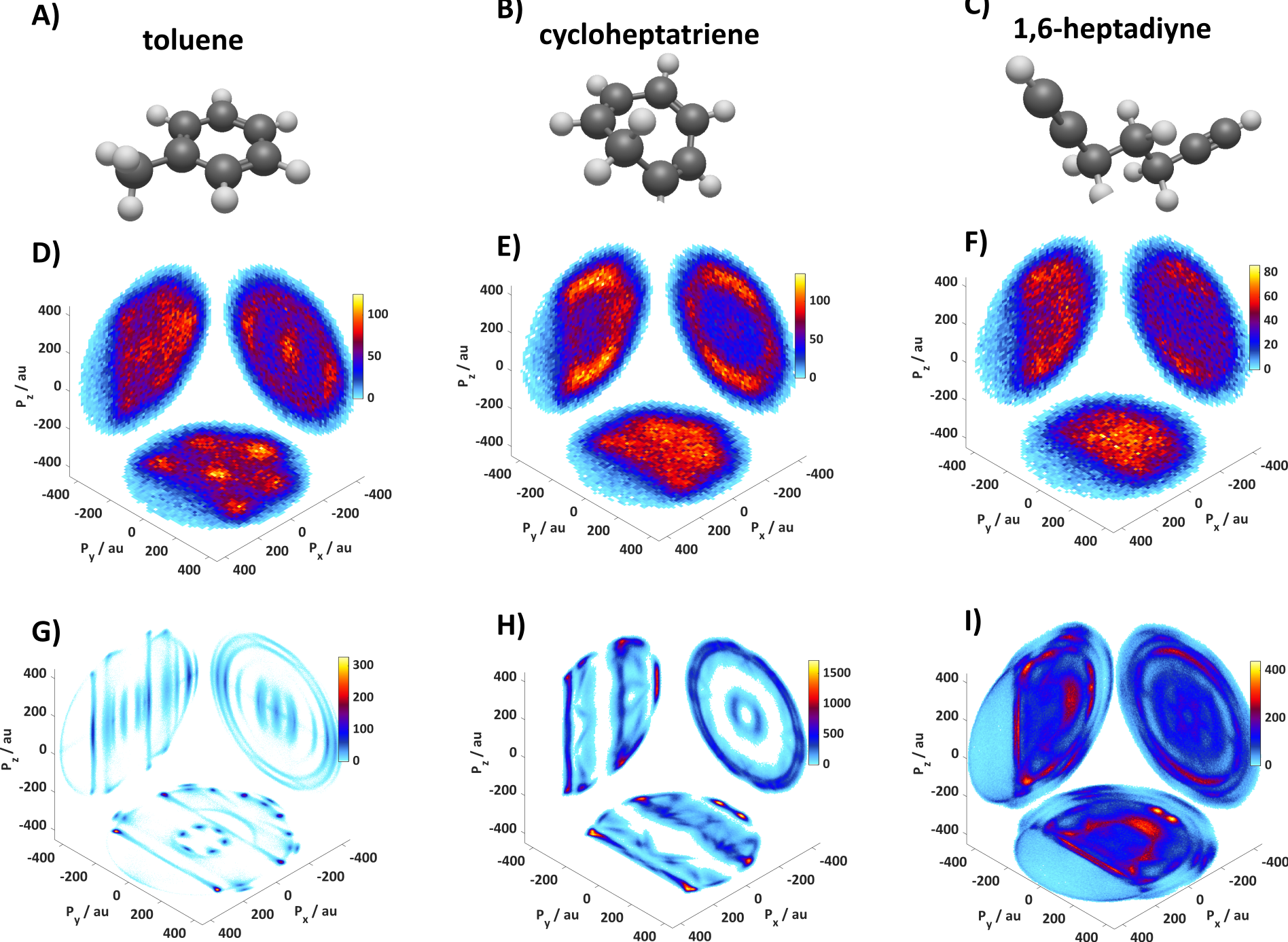}
        \caption{Molecular geometries and Newton plots: 
        A) toluene, B) cycloheptatriene, C) 1,6-heptadiyne. 
        D)-F) Experimental Coulomb explosion Newton plots showing the momentum distribution (in atomic units, a.u.) of three \ce{C^2+} ions detected in coincidence. The molecular frame is defined by orienting the emission direction of one \ce{C^2+} ion, which is randomly chosen among the three detected \ce{C^2+} ions, along the positive $p_x$-direction and the emission direction of another \ce{C^2+} ion, also randomly chosen from the three ions detected in coincidence, in the upper half of the $(p_x/p_y)$-plane. 
        G)-I) Simulated Newton plots for \ce{C^2+}, built by adding the simulated momentum distributions in all the recoil frames obtained from all possible combinations of two \ce{C^2+} ions, as explained in sections \ref{sec:sim} and \ref{sec:sim_results}. In the simulations, a charge of +2 is placed on each carbon atom and +1 on each hydrogen atom.}
        \label{fgr:newton_plot_main}
    \end{figure*}
Figure \ref{fgr:newton_plot_main} panels D) through F) show the experimental recoil-frame ion momentum distributions of doubly charged carbon ions obtained by detecting three \ce{C^2+} ions in coincidence after multi-photon X-ray ionization of D) toluene, E) cycloheptatriene, and F) 1,6-heptadiyne in the form of \textit{Newton plots}.         
Newton plots display the momentum distribution of each ion within a molecular \textit{recoil frame} defined by two selected ions. In this coordinate frame, one of the ion momenta ($\vec{p}_{i}$) defines the x-axis. Together with the momentum vector of a second ion ($\vec{p}_{j}$), it spans the \textit{recoil plane} incorporating the x-axis and the y-axis. The z-axis is defined as the normal to the recoil plane yielding a right-handed coordinate frame. The momenta of other measured fragments are transformed into this coordinate frame and plotted.
%with respect to this xy-plane.
%These breakup events are selected with three-particle coincidences.
For the plots shown here, the two reference ions as well as the third ion plotted in the figures were all selected to be \ce{C^2+}, and the two reference ions were randomly chosen from these three ions. We chose the \ce{C^2+}-\ce{C^2+}-\ce{C^2+} three-particle coincidence channel for the analysis because it is a relatively strong channel with sufficient statistics and because the charge of +2 on each ion implies that the molecule was ionized to a relatively high total charge state. The latter increases the likelihood that each atom in the molecule carried at least one charge during the explosion, which improves the quality of the Coulomb explosion images and facilitates comparison with our relatively simple model assuming purely Coulombic repulsion of classical point charges.
%, which is typically an increasingly better approximation the higher the total charge state. 
Similar Newton plots for toluene for other strong three-particle coincidence channels containing one or several \ce{C^+} fragments are shown in Fig.~6 in the Supplementary Material. %These are less clearly structured.
        % , suggesting that \ce{C^2+}-\ce{C^2+}-\ce{C^2+} channel is best suited for the purpose of identifying the molecular structure from the Coulomb explosion images.

%
    \begin{figure*}[t]
        \centering
        \includegraphics[height=12cm]{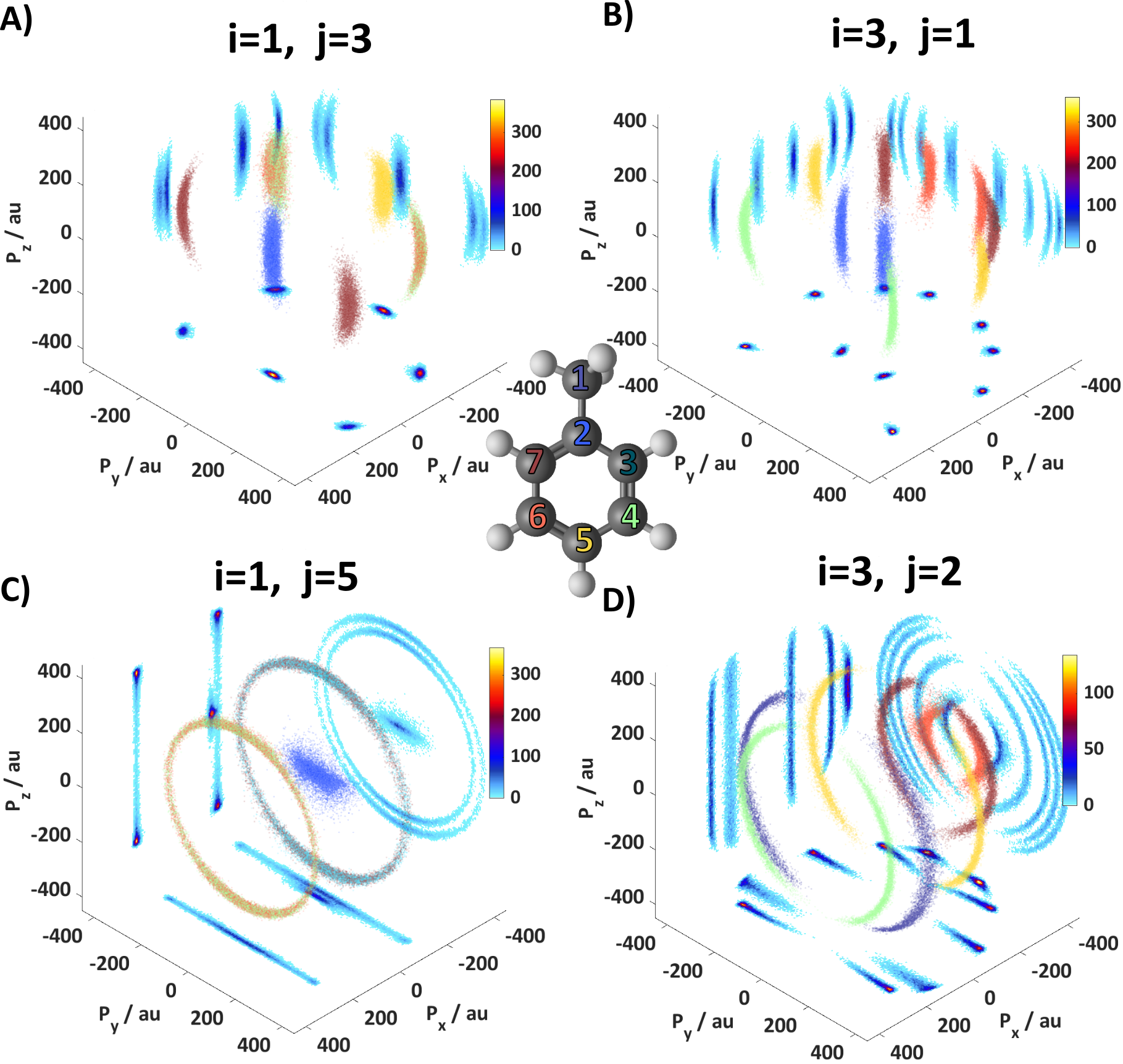}
        \caption{Simulated three-dimensional Newton plots of toluene, plotted for different combinations of reference ions, corresponding to different carbon sites in the molecule.
        Carbon \textit{i} denotes the ion momentum that defines the x-axis, while carbon \textit{j} is that which lies within the (x,y)-plane, with the numbering of the carbon atoms and the color coding of the scatter plot symbols referring to the ball-and-stick model of the molecule shown at the center of the figure. The momenta of the two reference ions are not shown in this representation, and projections of the momentum distributions on the three Cartesian planes are shown in light blue to aid with the three-dimensional visualization.}    
    \label{fgr:toluene_newton_reference_examples}
    \end{figure*}
As discussed above, in our previous CEI studies\cite{boll_x-ray_2022,pitzer_direct_2013, jahnke_direct_2025, lam_differentiating_2024, green_submitted}, the molecules under investigation 
%contained at least one 
consisted of at least two unique atomic constituents, allowing us to define the recoil frame for each molecule in a straightforward and intuitive way. The pure hydrocarbons studied here do not have such uniquely distinguishable atoms, which leads to an ambiguity in the assignment of the molecular frame based on the measured \ce{C^2+} ion momenta. For example, the molecule sketched in Fig.~\ref{fgr:newton_plot_main}A) is oriented with the methyl group pointing along the positive $p_x$-axis. However, as long as the reference ions are randomly chosen from the detected \ce{C^2+} ions and no other metrics are used to identify the emission site of a given ion, this specific orientation is only realized in a subset of events in which the carbon atom of the methyl group was (by chance) selected as the ion with momentum $\vec{p}_{i}$. In other cases, any of the other carbon ions maybe have been used to define the molecular-frame x-axis. The experimental Newton plots in Figs.~\ref{fgr:newton_plot_main}D)-F) are thus comprised of 42 different recoil frames (corresponding to the number of permutations of 2 items selected from a set of 7).
% , as illustrated by the simulations shown in Fig.\ \ref{fgr:toluene_newton_reference_examples}, which are discussed in more detail in the following.
We detail in sections \ref{sec:toluene} through \ref{sec:hep} how this problem can be circumvented to some extent.
        
Despite this summing over several different recoil frames, distinct differences between the three isomers can already be identified in Figs.~\ref{fgr:newton_plot_main}D)-F).
%, and the momentum distributions exhibit features that can be mapped to the equilibrium nuclear geometries, as we discuss in more detail in the following subsections.
The ion momenta of toluene (Fig.\ \ref{fgr:newton_plot_main}D) exhibit the most structured Newton plot 
%. The molecular breakup has a maximum at p$_z$ = 0 and results in 
with six distinct peaks in the $(p_x/p_y)$-plane. 
%(two of which, at positive p$_y$, appearing more faintly). 
In the $(p_y/p_z)$-plane, a low-momentum peak is surrounded by a distinct ring of higher momenta, but the distribution has a maximum at p$_z$ = 0.
The cycloheptatriene momentum distribution (Fig.\ \ref{fgr:newton_plot_main}E) does not have distinct peaks in the $(p_x/p_y)$-plane. However, it also shows a high-momentum ring in the $(p_x/p_z)$-plane, but without a clear maximum at p$_z$ = 0 and without the low-momentum peak. 
%shows a comparable ring in the xy-plane, but without a clear maximum around p$_z$ = 0. The low-momentum peak as well as the distinct peaks in the xy-plane are missing in this case. 
% has several broad features, none of which are clearly reminiscent of the molecular geometry .
% Most of the momentum is distributed in the yz- and zx-planes, signifying that fragments are ejected with a large momentum component perpendicular to the reference recoil plane.
Finally, the 1,6-heptadiyne Newton plot (Fig.\ \ref{fgr:newton_plot_main}F) shows a reminiscence of the ring in the $(p_y/p_z)$-plane, but significantly broadened compared to the other two molecules. Otherwise, the Newton plot shows a rather structureless distribution in all three planes. 
%We note that the non-cyclic 1,6-heptadiyne molecule can easily convert between different conformer geometries at room temperature (see also Fig.\ \ref{fgr:hep_good_ref}), which is not the case for the other isomers. 
%Several geometries of the parent molecules likely result in different contributions to the recoil frame (in addition to the common ambiguity of assigning the atomic site of the reference ions). Further insight into the origin of the observed features and their differences can be obtained from Coulomb explosion simulations, as discussed in the following section. 
%Note that the asymmetry along the $p_y$-axis in all of these Newton plots is an artifact of our recoil-frame definition. As the ion with momentum $\vec{p}_{j}$ defining the (x,y)-plane is not plotted, the corresponding intensity is missing here.  
    
\subsection{Simulation results}
\label{sec:sim_results}
In order to help with interpreting the experimental results, classical Coulomb explosion simulations are performed that can be used to assign the features observed in the experimental data. In the simulations, the momentum of each fragment ion under the influence of purely Coulombic interactions with the other ions is calculated in the center-of-mass frame, as described in section \ref{sec:sim}. The results are shown in Fig.\ \ref{fgr:newton_plot_main}, panels G)-I).
The momenta for each pair of two ions are used to construct the recoil frame, and the corresponding Newton plots for each possible recoil frame (see Fig.\ \ref{fgr:toluene_newton_reference_examples} for examples of some of the recoil frames) are combined to yield a simulated plot that can be directly compared to the experimental plots.
        
In the simulation, it is possible to specify which ions from certain atomic sites in the molecule are used as reference ions for the recoil frame.
        As expected, the features in the Newton plots vary significantly depending on the choice of reference ions.
        This observation is exemplified using toluene in Fig.\ \ref{fgr:toluene_newton_reference_examples}, where several simulated Newton plots with reference ions from different carbon sites in the molecule are shown.
        The momentum for each fragment ion is represented as color-coordinated points (allowing identification of the atomic site in the molecule), and the overall recoil-frame momentum distribution is projected onto the three planes (x,y), (x,z), and (y,z), in analogy to the representation in Fig.\ \ref{fgr:newton_plot_main}.
        %Just as for the experimental data, the reference momenta $\vec{p}_{i,j}$ are not shown.
        Some of these Newton plots are very reminiscent of the six-membered ring geometry of toluene, e.g., Fig.\ \ref{fgr:toluene_newton_reference_examples} panels A) and B), while in others, shown in Fig.\ \ref{fgr:toluene_newton_reference_examples} panels C) and D), obvious similarities are lost, as features become very diffuse in all three planes. The reason for this behavior can be seen, for example, in panel C), where the two ions defining the recoil frame are emitted close to back-to-back, thus not providing a well-defined recoil plane.   
        We therefore conclude that it would be beneficial for the determination of the molecular structure to try to select, as much as possible, which recoil frames contribute to the Newton plots of the experimental data such that unsuitable combination can be excluded, as opposed to including all combinations of reference pairs.
       %While the simulation allows for straightforward specification of reference ions and examination of the corresponding Newton plots, this is not as simple with the measured data.
       %The molecular structure of these hydrocarbons consists only of hydrogen and carbon atoms, preventing a unique recoil frame from being constructed by selecting fragments based solely on their mass-to-charge ratio. 
       Thus, in the following, we investigate ways to identify the molecular site from which specific ions originate in the experiment. Since the specifics depend on the particular molecular structure, we discuss each isomer separately.
    
    \subsection{Toluene}
    \label{sec:toluene}

        Toluene is a planar molecule consisting of a methyl group attached to a six-membered phenyl ring.
        Some features in the Newton plot of Coulomb-exploded toluene reflect its cyclic geometry, as shown in Fig.\ \ref{fgr:newton_plot_main} D and G.
        In the measured data, there are five distinct peaks in the $(p_x/p_y)$-plane, which are approximately equidistant from the origin, 
        %with a $|\bf{p}|$$\approx200~a.u.$ magnitude 
        and one peak lying near the origin of the $(p_x/p_y)$-plane.
       % these features appear as six broad features centered around $|\bf{p}|$$\approx200~a.u.$, 
        The five outer peaks correspond to the high-energy emission of the carbons on the phenyl ring (C3-C7, as labeled in Fig.~\ref{fgr:toluene_newton_reference_examples}) and the substituted methyl carbon (C1). The carbon to which the methyl group is attached (carbon C2) corresponds to the low-energy feature near the center of the $(p_x/p_y)$-plane because emission of carbon C2 is impeded due to opposing Coulomb repulsion from its neighboring ions at the methyl site and the rest of the carbon ring (also known as 'obstructed fragmentation'\cite{eland_dynamics_1987}). A similar behavior was observed in the Coulomb explosion of other cyclic molecules with a heavy substituent \cite{boll_x-ray_2022,green_submitted}. 
        
        Comparing the simulation results in Fig.\ \ref{fgr:newton_plot_main}G with the measured data, we find similarities but also distinct differences. First, the fact that the absolute momenta of the carbon atoms in the simulation (which are centered around $|\bf{p}|$$\approx400~a.u.$) are approximately twice the value of the experiment, where the outer peaks are centered around $|\bf{p}|$$\approx200~a.u.$, clearly indicates that the approximations made in the simulations are not suitable for accurately describing the absolute momentum magnitudes. Most notably, these assumptions are the instantaneous charge-up to the final charge state, the assumption of purely Coulombic potential energy surfaces of the highly charged ion, and also the assumption of a charge of 2+ for \emph{each} carbon and 1+ for \emph{each} hydrogen atom. In the experiment, the charges on the undetected ions are unknown and thus lead to a significant variation of the recoil momentum. 
        However, as also observed in earlier work \cite{boll_x-ray_2022, lam_differentiating_2024, jahnke_direct_2025, green_submitted}, despite overestimating the momentum magnitudes, such simulations typically reproduce the relative emission directions of the molecular fragments very well, albeit significantly underestimating the angular spread of each of the peaks. We can therefore use the simulation results to guide our attempts of better defining the recoil frame, as described in section \ref{sec:identification}. 
       In the simulated data, the aforementioned five outer peaks as well as the inner peak in the $(p_x/p_y)$-plane exhibit a clear sub-structure.
        This sub-structure in the simulated Newton plot results from the inclusion of all possible recoil frame combinations:
        The simulations in Fig.\ \ref{fgr:toluene_newton_reference_examples} show that Newton plots, where the carbon from the methyl group (C1) and a carbon other than C2 or C5 are chosen as references, provide an intuitive picture of the molecular geometry in the $(p_x/p_y)$-plane.
        %In the Newton plots the carbons on the phenyl ring to the high-momentum peaks on the outer rings and show an obstructed low-energy emission of carbon C2, as seen in Fig.\ \ref{fgr:toluene_newton_reference_examples}A. 
        However, when a different combination of reference ions is chosen, e.g., in Fig.\ 
%\ref{fgr:toluene_newton_reference_examples}A and Fig.\ 
       \ref{fgr:toluene_newton_reference_examples}B, all the peaks are slightly shifted, as can also be seen clearly in Fig.~7 in the Supplementary Material, which shows the Newton plots for all possible combinations of reference ions. These shifts are an additional reason why the experimental distributions are so broad, since in the experimental Newton plots, we integrated over all possible combinations of reference ions.

        In addition, if the reference ions are emitted anti-parallel, i.e., back-to-back, the reference ion $j$ will have no perpendicular component to reference ion $i$, and the $(p_x/p_y)$-plane in the recoil frame is thus ill-defined.
        This causes several features in the Newton plot to be dispersed along one of the axes in the projections, as shown in Fig.\ \ref{fgr:toluene_newton_reference_examples}C. Similarly, if the ion corresponding to the obstructed fragment (carbon C2) is included as one of the reference ions, the $(p_x/p_y)$-plane is also ill-defined, as this ion has a momentum close to $|\vec{p}|=0$~a.u. and thus an almost arbitrary emission direction compared to the other ions.
        In the recoil frame, this leads to strongly dispersed momentum distributions, especially outside the recoil plane, as shown in Fig.\ \ref{fgr:toluene_newton_reference_examples}D.
        %, and effective mapping to the molecular geometry is indirect.
        
\subsubsection{Identification of specific carbon ions via angular correlation metrics}
\label{sec:identification}

%The above simulations show that the superposition of different recoil frames due to the lack of unique marker atoms in pure hydrocarbons significantly complicates the visualization of the molecular geometry through Newton plots. Due to the planar geometry and high symmetry of toluene, some features linked to the molecular geometry can still be clearly identified in the Newton plot, such as the low-energy feature corresponding to the obstructed motion of carbon C2, but the other momentum peaks are difficult to resolve as several of the individual carbon atom signatures nearly overlap when all the recoil frame contributions are included.
The preceding subsection demonstrated the need for means to further identify experimentally which of the different carbon ions of the toluene molecule were detected in each triple-coincidence event in order to disentangle the different recoil frames that contribute to Fig.\ \ref{fgr:newton_plot_main}D).   
One advantage of momentum-resolved multi-particle coincidence measurements is that there are additional correlations that encode information on the molecular fragmentation, which can be used in particular for identifying suitable reference ions.
As we show in the following, events can be selected in which the momenta of the measured carbon ions meet certain criteria in their correlations that correspond to a specific carbon site (or subset of sites) inside the molecule.
Such a selection reduces the number of recoil frames that contribute to a Newton plot, thus providing a cleaner and more intuitive Coulomb explosion pattern.
For example, removing ions with low kinetic energy from the set of reference ions reduces the likelihood that the recoil plane is ill-defined because the reference ions do not include the carbon C2, which is undergoing obstructed breakup.
Moreover, the relative emission angle between two carbon ions provides information on the relative positions of these carbons in the molecule.

\begin{figure*}[t]
\centering
\includegraphics[height=10cm]{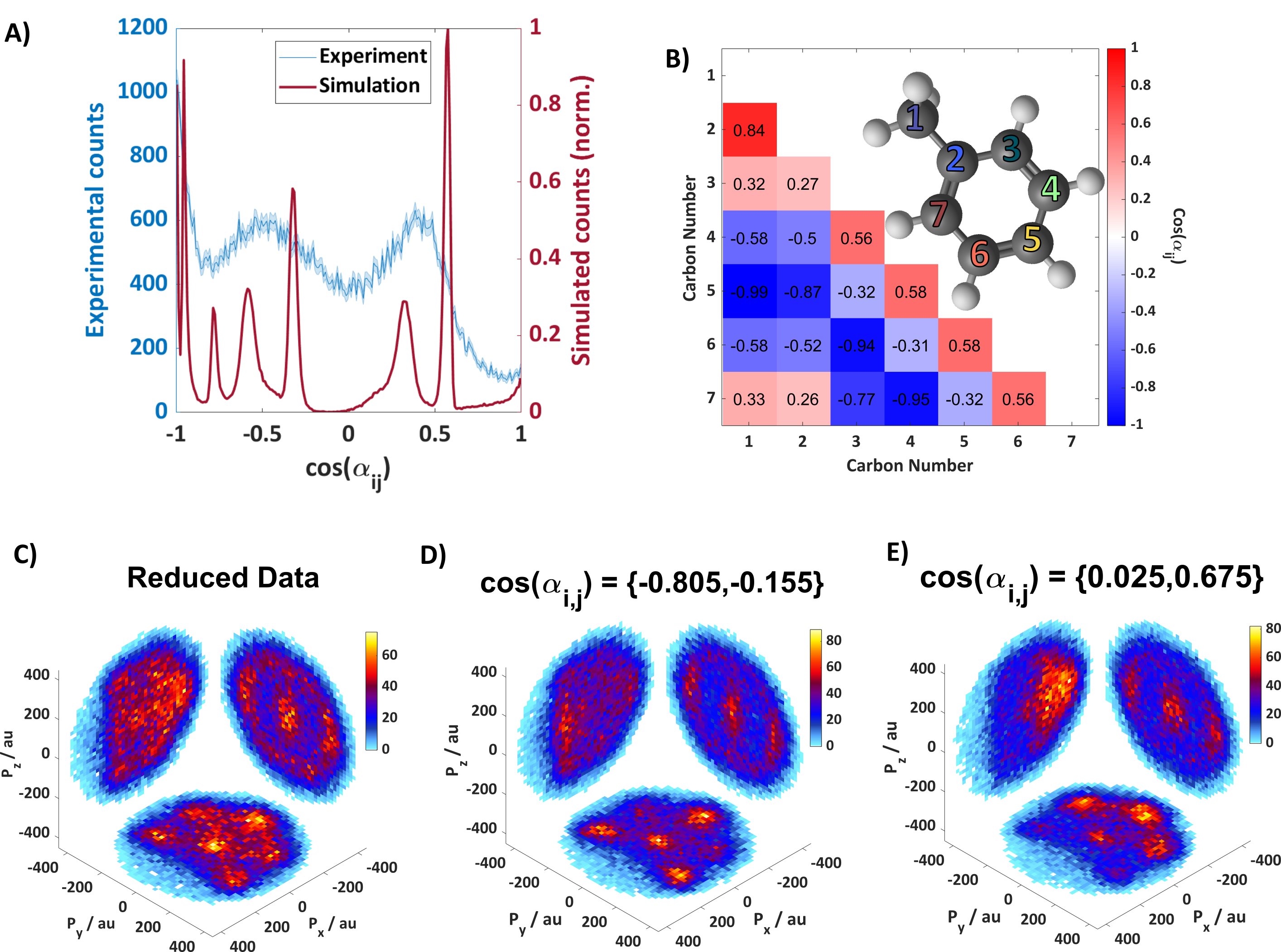}
\caption{
        Data filtering on angular correlations of the reference ions employed for defining the recoil frame of the toluene Newton plots.
        A) Simulated (red) and experimental (blue) distribution of the relative emission angle of the two ions $i$ and $j$ used to define the recoil frame.
        B) Matrix showing the average $\langle\cos(\theta_{i,j})\rangle$ in the simulated datasets for different combinations of carbon ions. 
        This matrix acts as a guide for which gating region corresponds to which reference ions defining the Newton plot.
        Darker blue colors correspond to $\langle\cos(\theta_{i,j})\rangle$ values closer to $-1$, darker red colors correspond to values closer to $+1$. 
        C) Toluene Newton plot without filtering.
        D), E) Toluene Newton plots with filtering on the relative angular distribution for the angular range indicated above each panel. 
        %Panel C is identical to panel D in Fig.\ \ref{fgr:newton_plot_main} but with some events randomly removed to have similar statistics as in panels D and E for better comparison. 
        }
\label{fgr:toluene_angular_distribution}
\end{figure*}

        A procedure for making use of some of this information is exemplified in Fig.\ \ref{fgr:toluene_angular_distribution}.
        Panel A) shows the distribution of relative emission angles between any two selected ions that are used as the reference ions $i$ and $j$ as obtained from both the experimental (blue) and simulated (red) data. As noted before, the features observed in the experiment are much broader than those obtained in the simulation but occur in the same angular ranges.
        Using the simulation data, we computed the average value $\langle\cos(\theta_{i,j})\rangle$ of these relative emission angles for different pairs of ions and summarized the resulting values in the matrix shown in Fig.~\ref{fgr:toluene_angular_distribution}B. The values are color-coded for illustrative purposes.
        This matrix reveals that the peak at $cos(\theta_{ij}) = - 1$ stems from pairs of carbon atoms located on opposite sides of the phenyl ring, while the smaller peak near $cos(\theta_{ij}) = + 1$ is due to the nearly parallel emission of the methyl carbon (C1) and carbon C2.
        The peaks near $cos(\theta_{ij}) \approx +0.4$ and $\approx -0.5$ correspond to neighboring and alternating carbons within the phenyl ring, respectively.
       
        Using the assignment obtained from the simulation, we can identify the contribution of ion pairs to the broad peaks in the measured angular distribution. Exploiting this information, it is possible to improve the clarity of the measured Newton plots by filtering on the relative emission angle between the reference ions. Panel C) shows once again the Newton plot without filtering, thus containing all reference combinations. Note that we removed a random subset of the data to obtain a similar amount of statistics as in panels D) and E) and thus facilitate a direct comparison. 
        The Newton plots shown in panels D) and E) contain only those events in which the reference ions are emitted at relative angles $cos(\theta_{ij}) \in \{-0.805, -0.155\}$ and $cos(\theta_{ij}) \in \{0.025, 0.675\}$, respectively.
        Additionally, these reference ions are selected to have a kinetic energy larger than 1 a.u., to avoid including the obstructed carbon C2 as a reference ion, which degrades the clarity of the Newton plots as mentioned above (see Figs.\ \ref{fgr:toluene_newton_reference_examples}C and \ref{fgr:toluene_newton_reference_examples}D).
        With this filtering, the Newton plots become more structured as the number of contributing recoil frames in the Newton plot is greatly reduced, from 42 to no more than sixteen possible reference combinations, and the reference ions are primarily pairs of alternating or neighboring carbons (in panels D) and E), respectively). This interpretation is confirmed by the notable reduced intensity of the peak on the positive and negative side of the $p_x$-axis, respectively. It would be in these empty spaces where the reference ion $j$ would appear, but this reference ion is not plotted here for the sake of demonstrating the efficiency of the filtering method. 

\subsection{Cycloheptatriene}
\label{sec:CHT}
\begin{figure*}[t]
\centering
\includegraphics[height=14cm]{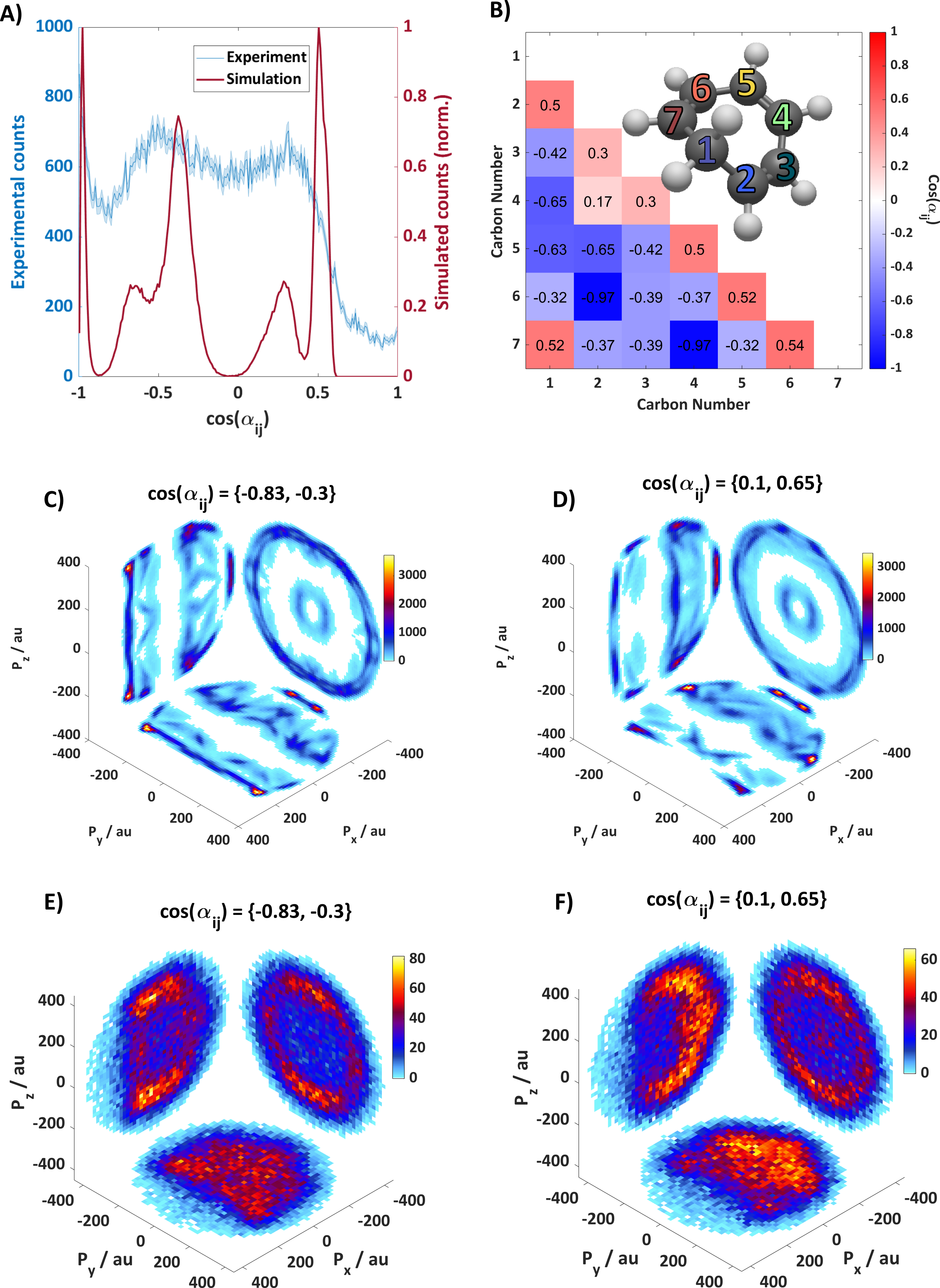}
\caption{
        Data filtering on angular correlations of the reference ions employed for defining the recoil frame of the cycloheptatriene Newton plots, similar to Fig.\ \ref{fgr:toluene_angular_distribution}, which shows the same for toluene.
        A) Simulated (red) and experimental (blue) distribution of the relative emission angle of the two ions $i$ and $j$ used to define the recoil frame.
        B) Matrix showing the average $\langle\cos(\theta_{i,j})\rangle$ in the simulated datasets for different combinations of carbon ions.
        C) and D) Simulated Newton plots filtered on the peaks in panel A) corresponding to alternating and neighboring carbons, respectively, selected by angular gating on the range of $\langle\cos(\theta_{i,j})\rangle=(-0.83, -0.3)$ and $(0.1, 0.65)$, respectively.
        E) and F) Corresponding experimental results for the same angular gating as in panels C) and D).
        }
\label{fgr:cyclo_angular_distribution}
\end{figure*}
Cycloheptatriene (CHT) is a seven-membered cyclic hydrocarbon. It has a non-planar chair conformation in its ground-state geometry. Because of the non-planarity, we expect a larger magnitude of the out-of-plane momenta
%a much stronger contribution of fragment momenta that do not lie within an assumed molecular plane 
than in the case of the toluene molecule. Indeed, this expectation is confirmed when inspecting the measured and simulated results of Coulomb-exploded CHT, which are shown in panels E) and H) of Fig.\ \ref{fgr:newton_plot_main}. The momentum distributions of the recoiling ions have strong contributions outside of the recoil plane defined by any combination of two reference ions and do not show a maximum at p$_z$ = 0. Furthermore, unlike the case of the planar toluene molecule, where the Newton plot had some resemblance to the molecular geometry, both the experimental and simulated Newton plots for CHT are much less intuitive.
%Symmetric features span only the yz and xz-planes in both the measured and simulated Newton plots. The xy-reference plane in the simulation contains non-symmetric features, while the measured data contains no discernible peaks, likely due to lack of resolution. In the yz-plane, distinct vertical stripes are present that appear as two co-centric circles when projected on the xz-plane. These features that extend out of the xy-plane are due to the non-planar properties of this molecule.
As for toluene, selecting reference ions based on their relative emission angles can reduce the number of contributing recoil frames, enabling easier mapping of features in the Newton plots to sites in the molecule.
The distribution of relative emission angles is shown in Fig.\ \ref{fgr:cyclo_angular_distribution}A) for both the simulated and experimental data. 
The large peak near $cos(\theta_{ij}) = -1$ is due to anti-parallel emission of carbon atoms located on opposite sides of the molecular ring.
Emission of ions from neighboring carbon sites (e.g., $i$ = 2 and $j$ = 3) corresponds to peaks near $cos(\theta_{ij}) \approx 0.5$, while ions from alternating carbon sites (e.g., $i$ = 2 and $j$ = 4) make up the peaks near $cos(\theta_{ij}) \approx -0.6$.
These features observed in the relative emission angle distribution for this seven-membered carbon ring are analogous to those belonging to the phenyl ring in toluene. We provide a matrix showing the average values of $\langle\cos(\theta_{i,j})\rangle$ for different pairs of ions in panel B) of Fig.\ \ref{fgr:cyclo_angular_distribution} just as in the previous subsection on toluene.  

\begin{figure*}[t]
\centering
\includegraphics[height=14cm]{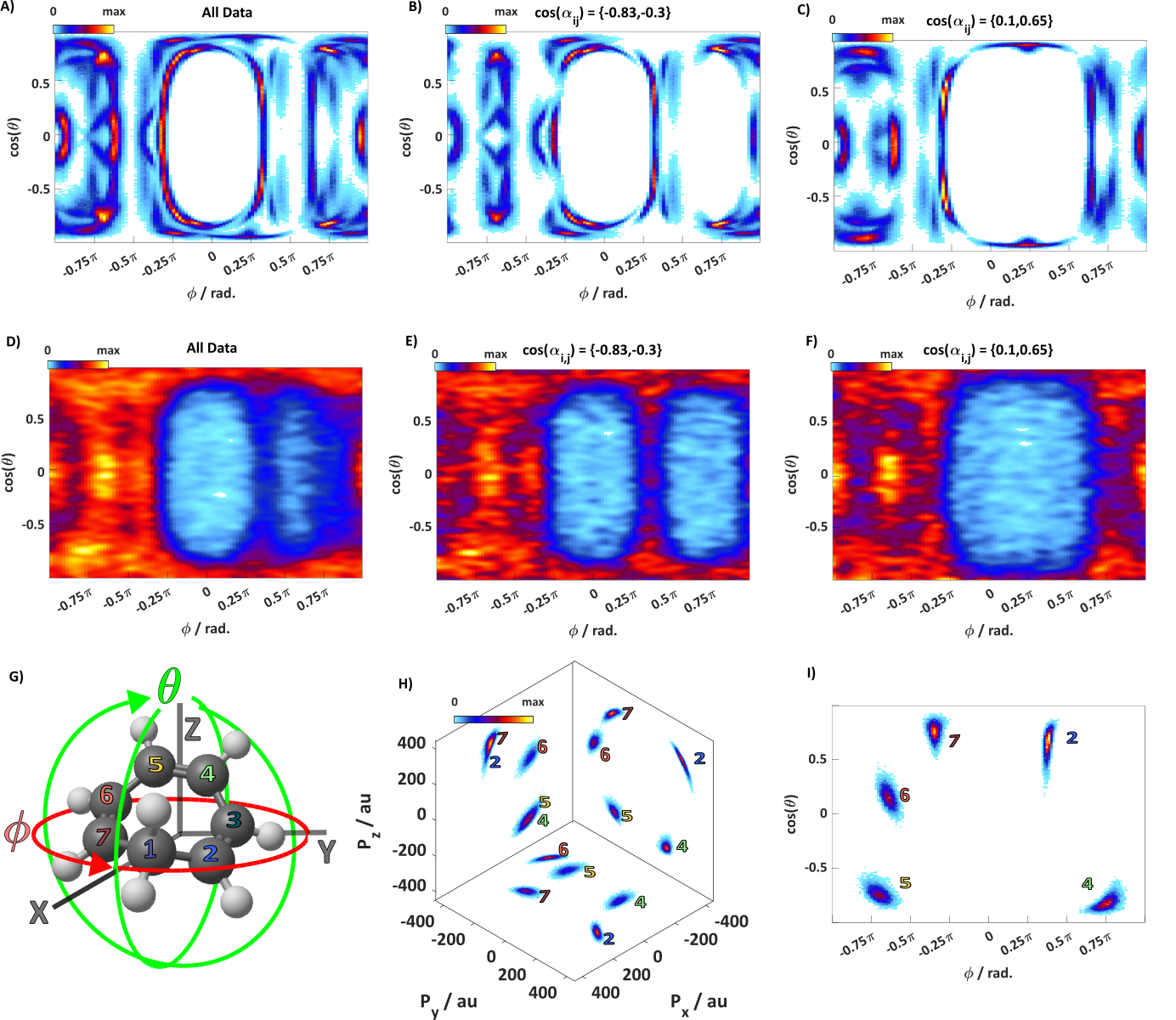} 
\caption{
            Molecular-frame angular emission distributions in spherical coordinates  for cycloheptatriene. The plots show the emission directions of ions (in the same molecular frame of reference as before) as a function of the polar ($\theta$) and azimuthal ($\phi$) angles for certain ranges of the relative angle $\alpha_{ij}$, similar to Fig.\ \ref{fgr:cyclo_angular_distribution}. 
            Simulated data are shown in the top row, panels A)-C), measured data in the middle row, panels D)-F). 
            G) Definition of the spherical coordinates.
            H) Simulated Newton plot for a recoil frame defined by only $i=1$ and $j=3$.
            H) Simulated molecular-frame angular distribution for this idealized recoil-frame definition.
        }
\label{fgr:cyclo_spherical_coordinates}
\end{figure*}
       Panels C) and E) of Fig.\ \ref{fgr:cyclo_angular_distribution} show the simulated and experimental Newton plots after gating on the aforementioned relative angle region associated with alternating carbons in the ring.
       %The momentum distribution along the x-axis lies on the positive side, as there is now the carbon between the reference ions that will emit nearly parallel with the references.
       The corresponding Newton plots for cases where preferably neighboring carbon ions were selected as reference ions $i,j$ are shown in panels D) and F). Comparing these two cases reveals three major differences. First, gating on the respective regions of relative emission directions indeed selects predominantly (even in the experiment) alternating or neighboring carbon ions in the molecule. The momentum of the first reference ion is located by definition at $p_x>0$~a.u., $p_y=0$~a.u, and $p_z=0$~a.u. Consequently, we do not observe intensity in the Newton plots in the region of $p_x>200$~a.u. Similarly, depending on the gate, distinct parts of the intensity in the $(p_x/p_y)$-plane are missing, which belong to either the neighboring carbon in panel F) or to one carbon next to the neighboring one in the alternating case in panel E). Furthermore, as expected for the non-planar chair conformation of the molecule, we observe more defined contributions in the $(p_x/p_z)$-plane at $p_z>200$~a.u.\ and $p_z<-200$~a.u., which is in line with an intuitive picture of the Coulomb explosion in a reference frame given by alternating carbons.    
       %For the case of neighboring carbons, as in panels D and F, a majority of the momentum distribution along the x-axis lies on the negative side, as the ions contributing to the Newton plot lie on the opposite sides of the carbon-ring framework respective of the neighboring reference carbons.
       The subtle peak at the center of the $(p_y/p_z)$-plane in Fig.\ \ref{fgr:cyclo_angular_distribution}D) and F) maps to the carbon atoms located in the same plane as the reference ions while emitted anti-parallel to one of them. 
       
\subsubsection{Molecular-frame angular emission distributions in spherical coordinates}

       Although the above discussion demonstrates that selecting events based on the reference ions' relative emission angles constructs different Newton plots, the differences between the reference selections are subtle in the case of CHT, and mapping the features to the carbon sites is difficult.  
       However, since all the emitted \ce{C^2+} ions from CHT are nearly equal in their kinetic energy, all the features in the Newton plot for this molecule lie on the surface of a spherical shell, and the data can thus be represented in form of molecular-frame angular emission distributions in spherical coordinates, with only minor loss of information when projecting over the radial coordinate. The same approach has previously been used to visualize the Coulomb explosion of the heterocyclic thiophenone molecule \cite{green_submitted} and to trace the time-dependent evolution of a thionucleobasis after UV excitation \cite{jahnke_direct_2025} using time-resolved CEI. 
       
       Figure \ref{fgr:cyclo_spherical_coordinates} shows such angular distributions in polar ($\theta$) and azimuthal angle ($\phi$) coordinates for the case of CHT. The simulated data are depicted in the top row (panels A-C) and the measured data in the middle row (panels D-F). 
       The definition of the spherical coordinates is illustrated in Fig.\ \ref{fgr:cyclo_spherical_coordinates}G.
       Although this representation typically does not provide an intuitive image of the molecular geometry, the dimensionality reduction of the fragmentation enables a more effective mapping of features to specific carbon sites in the molecule, as we demonstrate below.

 \begin{figure*}[t]
        \centering
        \includegraphics[height=8cm]{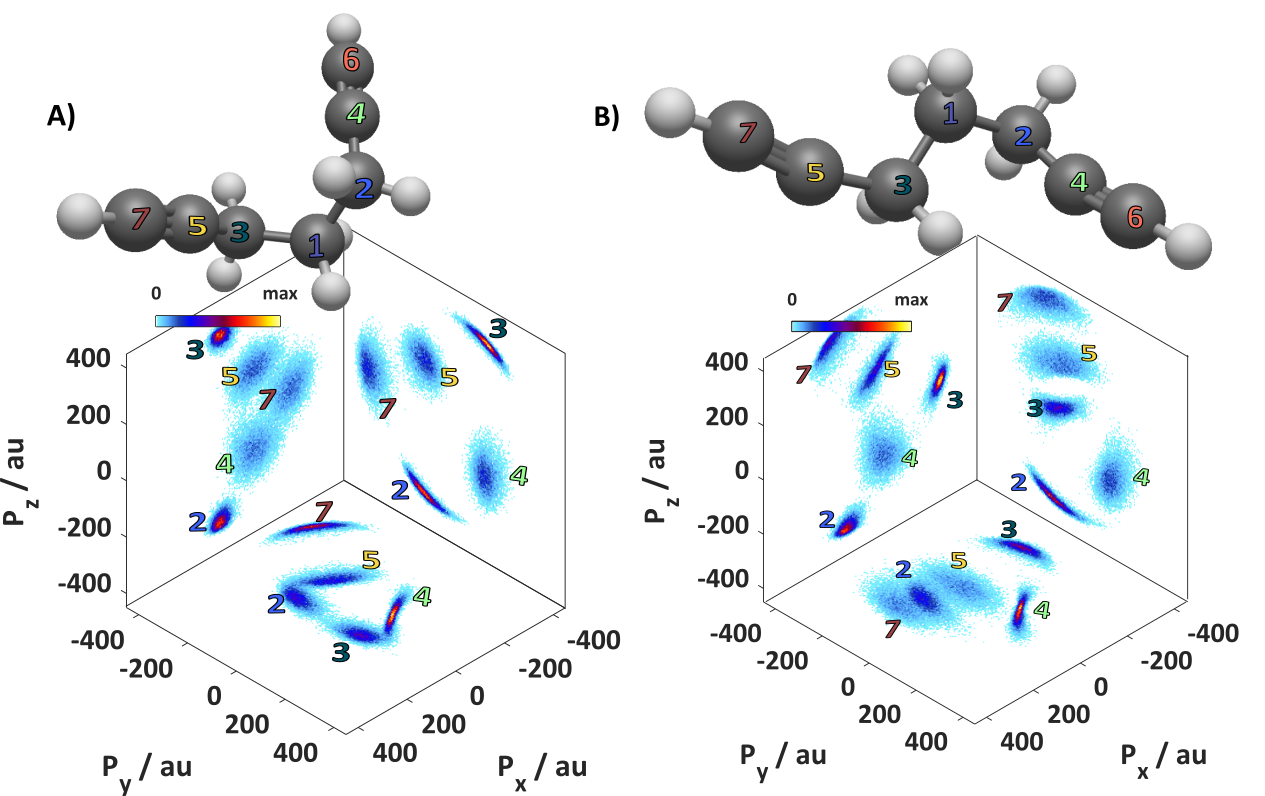}
        \caption{
        Two expected conformer geometries of the 1,6-heptadiyne molecule, (a) GG-trans and (b) AG, along with the corresponding simulated Newton plots plotted with carbons C1 and C6 as reference ions.
        }        
\label{fgr:hep_good_ref}
\end{figure*}  
       Panels A) and D) show the simulated and measured angular distributions without any filtering of the data. When filtering the data such that the reference ions are preferably given by alternating carbon atoms, we obtain the results shown in Figs.\ \ref{fgr:cyclo_spherical_coordinates}B) and \ref{fgr:cyclo_spherical_coordinates}E). A prominent band appears at $\phi \approx 0.38\pi$ in both experiment and simulation.
       This peak corresponds to the emission feature of the carbon atom located between the two reference ions (i.e., if $i=3$ and $j=5$, then this peak maps C4). For the case where neighboring carbon atoms are selected as reference ions, we obtain the molecular-frame angular distributions shown in Figs.\ \ref{fgr:cyclo_spherical_coordinates}C and \ref{fgr:cyclo_spherical_coordinates}F). There, the band near $\phi\approx 0.38 \pi$ occurs when the emitted atom is located near the reference ions, specifically the one closest to reference $j$ (e.g., if $i = 5$ and $j = 4$, this maps C6).   
       Lastly, signatures of the emission from the opposite side of the ring relative to the reference ions map to the peaks near $\phi \approx -0.63 \pi$ and $\phi \approx \pi$, depending on the $i,j$ reference permutation. For example, if $i = 3$ and $j = 4$ are chosen, then the peak occurs at $\phi \approx -0.63\pi$ and corresponds to carbon C7. When flipping the reference ions to $i = 4$ and $j = 3$, this carbon site is mapped to an angle of $\phi \approx \pi$.

       For comparison, Figs.\ \ref{fgr:cyclo_spherical_coordinates}H and \ref{fgr:cyclo_spherical_coordinates}I show the simulated Newton plots and angular emission distributions, respectively, for an idealized recoil frame defined by carbons C1 and C3, as illustrated in Fig.\ \ref{fgr:cyclo_spherical_coordinates}G), which represents one of the contributions to panels B) and E). Panel I) is asymmetric in $\cos(\theta)$, while the distributions shown in panels B) and E) are symmetric. A symmetric distribution in panel I) would occur if, in addition, the results for reference ions $i=1$ and $j=6$ (which corresponds to a rotation of the molecule sketched in panel G) around the $p_x$-axis) were superimposed. With this in mind, the measured and simulated angular distributions shown in panels E) and B) become intuitively understandable, and different regions can indeed be assigned to certain carbon sites in the molecule.       

\subsection{1,6-heptadiyne}
  \label{sec:hep}
       
Finally, for the non-cyclic isomer 1,6-heptadiyne, the single bonds that bridge different groups within the molecule allow for an easy interconversion between different conformer geometries, with two conformers, GG-trans and AG, expected to be dominant.\cite{restrepo_alkyl_2007}
%This interconversion can be induced thermally, as the barrier between these different configurations is low, typically on the order of $\sim200$ meV for these types of hydrocarbons. Even at a sample temperature as low as 2 K, which is significantly lower than the molecular beam temperature in this experiment, two primary conformers are expected to be present: GG-trans ($\sim81\%$) and AG ($\sim16\%$), with all other conformer configurations being negligible.\cite{restrepo_alkyl_2007} 
The geometries of these two primary conformers are shown in Fig.\ \ref{fgr:hep_good_ref} together with simulated Newton plots that were generated using ions $i=1$ and $j=6$ as references for the coordinate frame. 
%Here, there is no symmetry in the xy- plane. The xz - and yz - plane contains significant mirror symmetry about the x and y- axis, respectively, with all features off axis.
The two Newton plots demonstrate that the simulated momentum distributions of the GG-trans and AG conformers look distinctly different when constructing the Newton plots using a well-defined reference frame. However, they still do not directly map to an intuitive picture of the molecular geometry. Comparing these results to the case of having arbitrary reference ions, as shown in Fig.\ \ref{fgr:newton_plot_main}I), and, in particular, to the corresponding measured results in Fig.\ \ref{fgr:newton_plot_main}F) indicates that in the case of 1,6-heptadiene, further information present in the CEI data need to be evaluated in order to discriminate molecular structures. Thus, in the following subsection, we inspect the measured proton momenta with the goal of using them as messengers of the molecular structure, as successfully applied in our recent work on the Coulomb explosion of thiophenone and thiouracil \cite{jahnke_direct_2025, green_submitted}.  
        %As the experimental data containing multiple molecular reference frames are unable to resolve the fine substructure in the simulated CEI patterns, exemplified also by the results of the previous two isomers, it is not unexpected that the experimental Newton plot for 1,6-heptadiyne, shown in Fig.~\ref{fgr:newton_plot_main}F, is rather unstructured in all three planes, making it impossible to draw any conclusions, e.g., regarding the ratio of the different conformers contributing to the data.
        %likely due to insufficient experimental resolution and deviations from the assumed prompt charge up and explosion process, we might expect this with the several fine structures in the Newton plot for the conformers.
        
        %The reason for the discrepancy between experimental and simulated Newton plots, beyond the limited resolution, could be due to several other conformer geometries that are probed. 
        %Figure \ref{fgr:newton_plot_main}I includes the simulated results of Coulomb exploding an admixture of the two different conformers of 1,6-heptadiyne shown in Fig.\ \ref{fgr:hep_good_ref}. 
        %, due to the non-planar properties of this GG-trans conformation geometry.

\subsection{Molecular-frame angular emission distributions of protons}
\label{sec:hydrogens}
 \begin{figure*}[t]
        \centering
        \includegraphics[height=8cm]{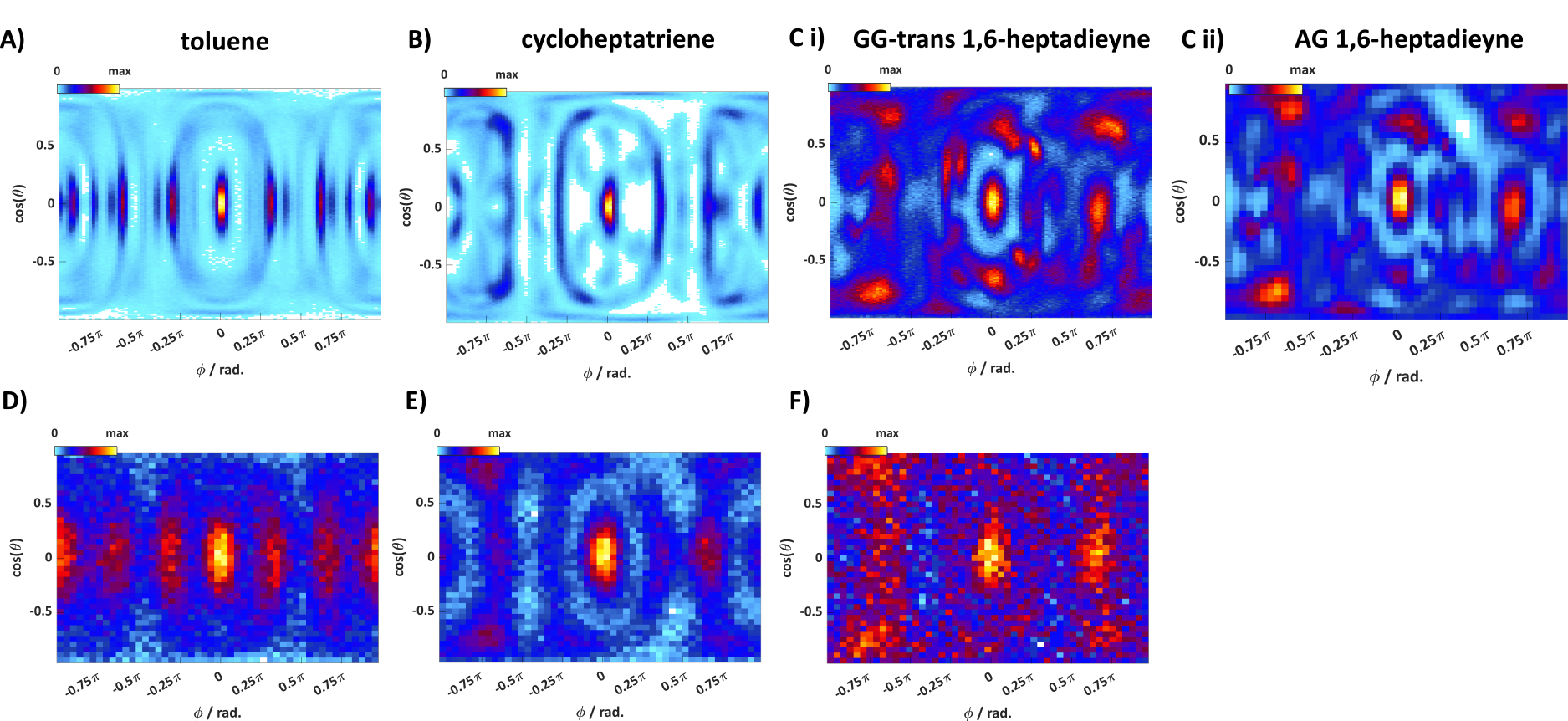}
        \caption{
         Molecular-frame angular distributions of the emitted \ce{H+} ions obtained from the \ce{H+} + \ce{C^2+} + \ce{C^2+} coincidence channel, with the two \ce{C^2+} ions defining the recoil frame. A)-C): Simulation and D)-F) experimental data for toluene, cycloheptatriene, and 1,6-heptadiyne, respectively. Simulations for both the GG-trans and the AG conformers of 1,6-heptadiyne are shown in panels C i) and C ii), respectively.  
        }
        \label{fgr:hydrogen_imaging}
    \end{figure*}
        We have shown above how Newton plots of the carbon atoms can be used to distinguish between the different \ce{C7H8} isomers and how their relative momentum distributions in the recoil frame can provide information on the geometric properties of these hydrocarbon molecules. We now turn to the H$^+$ ion momentum distributions and show that they also contain rich structural information, which complements the information encoded in the carbon ions and, when combined, provides additional metrics for distinguishing the structures.
        The experimental and simulated molecular-frame angular distributions of the \ce{H^+} fragments are shown in Fig.\ \ref{fgr:hydrogen_imaging} for all three isomers, toluene, cycloheptatriene, and 1,6-heptadiyne.
        These plots are made by selecting the three-particle coincidence channel \ce{H^+}+\ce{C^2+}+\ce{C^2+} and using the two \ce{C^2+} fragments to define the recoil frame in the same way as for the corresponding plots in the previous figures. The corresponding 3D Newton plots for the \ce{H^+}, \ce{C^2+}, \ce{C^2+} channel for all three isomers are shown in Fig.~5 of the Supplementary Material. 
        
        For the case of toluene, the majority of the protons are emitted in the plane spanned by the momenta of the \ce{C^2+}+\ce{C^2+} reference ions (as evidenced by the peaks centered around $\cos{\theta}\approx0$). 
        The corresponding plots for cycloheptatriene in panels B) and D) demonstrate that for this nonplanar molecule, the \ce{H^+} ion emission also has an additional, strong out-of-plane component located near $\cos(\theta)\approx \pm1$. %The out-of-plane emission is even 
        Such contributions far from $\cos{\theta}\approx0$ are even more pronounced for 1,6-heptadiyne, as shown in panels C) and E). For the sake of clarity, the simulated angular distributions for both the GG-trans and AG conformers of 1,6-heptadiyne, corresponding to the geometries sketched in Fig.\ \ref{fgr:hep_good_ref}, are shown. 
        
        Note that in the angular distributions for all isomers, the bright spot at the center of the plot corresponds to the \ce{H^+} originating from the hydrogen connected to one of the reference carbons.

        The visually clear differences between the molecular-frame angular distributions of H$^+$ ions for all three isomers demonstrate that the H$^+$ ion momentum distributions are particularly well suited to distinguish isomers, as also shown recently for the case of the heterocyclic molecule thiophenone \cite{green_submitted}, and that changes in the H$^+$ emission pattern can thus be used as reporters for subtle changes in the carbon backbone of the molecule \cite{jahnke_direct_2025}.

\section{Conclusions}
    We have investigated the prospects of X-ray-induced CEI of pure hydrocarbons, in which there is no preferential absorption site, nor can a unique recoil frame be defined for the interpretation of the CEI data, as there are no unique atomic sites in the molecule that can act as references for the momenta of other ions. Despite the absence of such unique atomic markers, the measured momentum correlation patterns show discernible differences among the three isomers. For example, the toluene breakup yields a more structured Newton plot with multiple high-momentum ions arranged in a ring-like pattern reflecting the planar ring geometry, and a low-momentum feature near the origin, which can be attributed to the carbon nearest to the methyl group, whose emission is obstructed by the opposing repulsion from the neighboring carbon atoms. The cycloheptatriene momentum distribution is broader and less obviously reflective of a ring, with significant out-of-plane momentum components, consistent with the non-planar “chair” geometry of CHT. Finally, 1,6-heptadiyne, which can adopt many conformations, produces an almost homogeneous and isotropic carbon ion momentum distribution as a result of the 'floppiness' of the molecule. 
        
    Furthermore, by examining hydrogen ion (\ce{H+}) momentum distributions in coincidence with two \ce{C2+} ions, we observe that in the case of toluene, the \ce{H+} ions are preferentially ejected within the plane of the benzene ring, whereas for cycloheptatriene and especially 1,6-heptadiyne, the \ce{H+} ion momentum distributions exhibit much stronger components out of the plane spanned by the two reference ions – consistent with the non-planarity and conformational flexibility of those molecules.
    
    By comparison to classical Coulomb explosion simulations, we can assign specific features in the Newton plots and molecular-frame angular emission distributions to specific molecular sites. For example, in toluene, the low-momentum central peak corresponds to the ring carbon bonded to the CH$_3$ group (experiencing opposing Coulomb forces from both the methyl group and the rest of the ring), whereas the outer-ring momentum peaks correspond to carbons on the benzene ring and the methyl carbon itself. By filtering events based on the relative angle between two \ce{C2+} fragment momenta and excluding low-kinetic-energy reference ions, we can isolate subsets of data associated with particular pairs of originating carbon sites. This filtering simplifies and 'cleans up' the momentum maps by reducing the contributing reference-frame combinations, making the correspondence to the molecular structure more evident.

    Since the selection of these subsets degrades the statistics of the resulting Newton plots, the next generation of high-repetition-rate sources, both table-top and XFELs, will significantly improve the applicability of this selection method. We also expect %the experimental resolution, both in energy and in angle,
    the clarity of the correlation plots to improve as higher-order coincidence events can be analyzed, which reduces the uncertainty regarding the overall charge state of the molecule that leads, in parts, to the broad width of the measured energy and angular distributions.
    
    Another avenue to reduce the ambiguity of the molecular reference frames for certain suitable molecules, such as toluene, could be isotope labeling, either by deuterium tagging or by replacing one of the carbon atoms with a carbon-13 isotope. However, depending on the spectrometer geometry and extraction voltages used for the experiment, it is typically not straightforward to separate highly energetic carbon ions with only one atomic mass unit difference as their time-of-flight peaks strongly overlap. This problem is alleviated in the case of 'complete' coincidences, i.e., for events where \textit{all} fragment ions are detected, because momentum conservation between the emitted ions leads to narrow coincidence lines that allow for high mass resolution. Currently, the highest-order multi-ion coincidence momentum imaging results that we are aware of are for 8-ion coincidences recorded with a 3-kHz femtosecond laser\cite{venkatachalam2025exploiting}, but high-repetition-rate XFELs and high-power table-top femtosecond laser sources with repetition rates of 100 kHz or more are now available, suggesting that 'complete' coincidences may be achievable even for molecules with more than 10 atomic constituents.
    
    Overall, the present study demonstrates that even without heavy atoms and unique markers, multi-ion coincidence CEI can still provide site-specific signatures of atomic positions (e.g., identifying which carbon was obstructed in toluene, or differences in hydrogen emission patterns), thereby distinguishing structural isomers. In future studies, machine learning tools and neural networks can be applied to the analysis of CEI data to help differentiating isomers formed as competing reaction products in a photochemical reaction \cite{venkatachalam2025exploiting} and reconstructing their real-space molecular structure \cite{li_AI_submitted, Ghanaatian_AI_submitted}. 
New technical developments with regard to both experimental capabilities and data analysis therefore continue to lead to much improved capabilities, making CEI a promising candidate for time-resolved studies of ultrafast structural rearrangements (such as isomerization reactions) in pure hydrocarbons.

\section*{Author Contributions}
AR and RB conceived the experiment, following initial discussions with DRo, MC, and PMW. The experiment was carried out by KDB, RB, TMB, SB, KC, BE, ADF, MI, EK, HVSL, XL, TMa, MM, TMu, SP, DRi, PS, FT, SU, ASV, EW, TJ, and DRo, with remote participation of MC, RF, LM, JPFN, AO, PMW, and AR. KDB analyzed the data, with help and guidance from RB, TJ, AR, and DRo. KB, RB, TJ, AR, and DRo interpreted the results and wrote the manuscript with input from all authors.
% We strongly encourage authors to include author contributions and recommend using \href{https://casrai.org/credit/}{CRediT} for standardised contribution descriptions. Please refer to our general \href{https://www.rsc.org/journals-books-databases/journal-authors-reviewers/author-responsibilities/}{author guidelines} for more information about authorship.
    % commented out by KB 21-01-2025

\section*{Conflicts of Interest}
There are no conflicts to declare.

\section*{Data Availability}
Data recorded for the experiment at the European XFEL are available at https://doi.org/10.22003/XFEL.EU-DATA-002448-00.

% A data availability statement (DAS) is required to be submitted alongside all articles. Please read our \href{https://www.rsc.org/journals-books-databases/author-and-reviewer-hub/authors-information/prepare-and-format/data-sharing/#dataavailabilitystatements}{full guidance on data availability statements} for more details and examples of suitable statements you can use.
    % commented out by KB 21-01-2025

\section*{Acknowledgements}
We acknowledge the European XFEL in Schenefeld, Germany, for the provision of X-ray free-electron laser beamtime at the SQS instrument through proposal no.~2448 and thank the EuXFEL staff for their assistance. This work was primarily supported by the Chemical Sciences, Geosciences, and Biosciences Division, Office of Basic Energy Sciences, Office of Science, US Department of Energy, who funded KDB, MC, KC, AO, AR, and DRo through grant no.~DE-SC0020276, and HVSL, SP, and EW through grant no.~DE-FG02-86ER13491. XL and RF were supported by the Linac Coherent Light Source (LCLS), SLAC National Accelerator Laboratory, which is funded by the U.S.~Department of Energy, Office of Science, Office of Basic Energy Sciences under Contract No.~DE-AC02-76SF00515. ASV was supported by the National Science Foundation grant no.~PHYS-2409365. PMW and LM acknowledge support by the National Science Foundation grant no.~CHE-2309434. FT acknowledges funding by the Deutsche Forschungsgemeinschaft (DFG, German Research  Foundation) - Project 509471550, Emmy Noether Program. MI was partly supported by the Bundesministerium für Bildung und Forschung (BMBF) under grant 13K22CHA. MI and MM acknowledges support by the Cluster of Excellence ‘Advanced Imaging of Matter’ of the DFG—EXC 2056 and project ID 390715994.

%%%END OF MAIN TEXT%%%


\balance


%%%REFERENCES%%%
\bibliography{rsc} %You need to replace "rsc" on this line with the name of your .bib file

@article{stapelfeldt_time-resolved_1998,
	title = {Time-resolved {Coulomb} explosion imaging: {A} method to measure structure and dynamics of molecular nuclear wave packets},
	volume = {58},
	issn = {1050-2947, 1094-1622},
	shorttitle = {Time-resolved {Coulomb} explosion imaging},
	url = {https://link.aps.org/doi/10.1103/PhysRevA.58.426},
	doi = {10.1103/PhysRevA.58.426},
	language = {en},
	number = {1},
	urldate = {2019-07-08},
	journal = {Physical Review A},
	author = {Stapelfeldt, Henrik and Constant, Eric and Sakai, Hirofumi and Corkum, Paul B.},
	month = jul,
	year = {1998},
	pages = {426--433},
}

@article{green_submitted,
	title = {Visualizing the three-dimensional arrangement of hydrogen atoms in organic molecules by {Coulomb} explosion imaging},
	author = {Green, A.E. and Chen, K. and Bhattacharyya, S. and Allum, F. and Usenko, S. and Ashfold, M. and Baumann, T. and Borne, K. and Brouard, M. and Burt, M. and Curchod, B. and Erk, B. and Forbes, R. and Ibele, L. and Ingle, R. and Lam, H.V.S. and Li, X. and Lin, K. and Mazza, T. and McManus, J. and Meyer, M. and Mullins, T. and Figueira Nunes, J.P. and Rivas, D. and Roerig, A. and Rouzee, A. and Schmidt, P. and Searles, J. and Senfftleben, B. and Stapelfeldt, H. and Tanyag, R.M. and Trinter, F. and Venkatachalam, A.S. and Wang, E. and Warne, E. and Weber, P.M. and Wolf, T.J.A. and Jahnke, T. and Rudenko, A. and Boll, R. and Rolles, D.},
	journal = {Journal of the American Chemical Society},
    volume = {147},
    pages = {37133-37143},
	year = {2025},
}

@article{hansen_torsion_2012,
    author = {Hansen, Jonas L. and Nielsen, Jens H. and Madsen, Christian Bruun and Lindhardt, Anders Thyboe and Johansson, Mikael P. and Skrydstrup, Troels and Madsen, Lars Bojer and Stapelfeldt, Henrik},
    title = {Control and femtosecond time-resolved imaging of torsion in a chiral molecule},
    journal = {The Journal of Chemical Physics},
    volume = {136},
    number = {20},
    pages = {204310},
    year = {2012},
    month = {05},
    issn = {0021-9606},
    doi = {10.1063/1.4719816},
    url = {https://doi.org/10.1063/1.4719816},
    eprint = {https://pubs.aip.org/aip/jcp/article-pdf/doi/10.1063/1.4719816/13889168/204310_1_online.pdf},
}

@article{jahnke_direct_2025,
	title = {Direct observation of ultrafast symmetry reduction during internal conversion of 2-thiouracil using {Coulomb} explosion imaging},
	volume = {16},
	copyright = {2025 The Author(s)},
	issn = {2041-1723},
	url = {https://www.nature.com/articles/s41467-025-57083-3},
	doi = {10.1038/s41467-025-57083-3},
	number = {1},
	urldate = {2025-03-03},
	journal = {Nature Communications},
	author = {Jahnke, Till and Mai, Sebastian and Bhattacharyya, Surjendu and Chen, Keyu and Boll, Rebecca and Castellani, Maria Elena and Dold, Simon and Frühling, Ulrike and Green, Alice E. and Ilchen, Markus and Ingle, Rebecca and Kastirke, Gregor and Lam, Huynh Van Sa and Lever, Fabiano and Mayer, Dennis and Mazza, Tommaso and Mullins, Terence and Ovcharenko, Yevheniy and Senfftleben, Björn and Trinter, Florian and Atia-Tul-Noor and Usenko, Sergey and Venkatachalam, Anbu Selvam and Rudenko, Artem and Rolles, Daniel and Meyer, Michael and Ibrahim, Heide and Gühr, Markus},
	month = feb,
	year = {2025},
	pages = {2074},
}

@article{jahnke_inner-shell-ionization-induced_2021,
	title = {Inner-{Shell}-{Ionization}-{Induced} {Femtosecond} {Structural} {Dynamics} of {Water} {Molecules} {Imaged} at an {X}-{Ray} {Free}-{Electron} {Laser}},
	volume = {11},
	url = {https://link.aps.org/doi/10.1103/PhysRevX.11.041044},
	doi = {10.1103/PhysRevX.11.041044},
	number = {4},
	urldate = {2021-12-06},
	journal = {Physical Review X},
	author = {Jahnke, T. and Guillemin, R. and Inhester, L. and Son, S.-K. and Kastirke, G. and Ilchen, M. and Rist, J. and Trabert, D. and Melzer, N. and Anders, N. and Mazza, T. and Boll, R. and De Fanis, A. and Music, V. and Weber, Th. and Weller, M. and Eckart, S. and Fehre, K. and Grundmann, S. and Hartung, A. and Hofmann, M. and Janke, C. and Kircher, M. and Nalin, G. and Pier, A. and Siebert, J. and Strenger, N. and Vela-Perez, I. and Baumann, T. M. and Grychtol, P. and Montano, J. and Ovcharenko, Y. and Rennhack, N. and Rivas, D. E. and Wagner, R. and Ziolkowski, P. and Schmidt, P. and Marchenko, T. and Travnikova, O. and Journel, L. and Ismail, I. and Kukk, E. and Niskanen, J. and Trinter, F. and Vozzi, C. and Devetta, M. and Stagira, S. and Gisselbrecht, M. and Jäger, A. L. and Li, X. and Malakar, Y. and Martins, M. and Feifel, R. and Schmidt, L. Ph. H. and Czasch, A. and Sansone, G. and Rolles, D. and Rudenko, A. and Moshammer, R. and Dörner, R. and Meyer, M. and Pfeifer, T. and Schöffler, M. S. and Santra, R. and Simon, M. and Piancastelli, M. N.},
	month = dec,
	year = {2021},
	pages = {041044},
}

@article{decking_mhz-repetition-rate_2020,
	title = {A {MHz}-repetition-rate hard {X}-ray free-electron laser driven by a superconducting linear accelerator},
	volume = {14},
	copyright = {2020 The Author(s), under exclusive licence to Springer Nature Limited},
	issn = {1749-4893},
	url = {https://www.nature.com/articles/s41566-020-0607-z},
	doi = {10.1038/s41566-020-0607-z},
	number = {6},
	urldate = {2020-07-30},
	journal = {Nature Photonics},
	author = {Decking, W. and Abeghyan, S. and Abramian, P. and Abramsky, A. and Aguirre, A. and Albrecht, C. and Alou, P. and Altarelli, M. and Altmann, P. and Amyan, K. and Anashin, V. and Apostolov, E. and Appel, K. and Auguste, D. and Ayvazyan, V. and Baark, S. and Babies, F. and Baboi, N. and Bak, P. and Balandin, V. and Baldinger, R. and Baranasic, B. and Barbanotti, S. and Belikov, O. and Belokurov, V. and Belova, L. and Belyakov, V. and Berry, S. and Bertucci, M. and Beutner, B. and Block, A. and Blöcher, M. and Böckmann, T. and Bohm, C. and Böhnert, M. and Bondar, V. and Bondarchuk, E. and Bonezzi, M. and Borowiec, P. and Bösch, C. and Bösenberg, U. and Bosotti, A. and Böspflug, R. and Bousonville, M. and Boyd, E. and Bozhko, Y. and Brand, A. and Branlard, J. and Briechle, S. and Brinker, F. and Brinker, S. and Brinkmann, R. and Brockhauser, S. and Brovko, O. and Brück, H. and Brüdgam, A. and Butkowski, L. and Büttner, T. and Calero, J. and Castro-Carballo, E. and Cattalanotto, G. and Charrier, J. and Chen, J. and Cherepenko, A. and Cheskidov, V. and Chiodini, M. and Chong, A. and Choroba, S. and Chorowski, M. and Churanov, D. and Cichalewski, W. and Clausen, M. and Clement, W. and Cloué, C. and Cobos, J. A. and Coppola, N. and Cunis, S. and Czuba, K. and Czwalinna, M. and D’Almagne, B. and Dammann, J. and Danared, H. and de Zubiaurre Wagner, A. and Delfs, A. and Delfs, T. and Dietrich, F. and Dietrich, T. and Dohlus, M. and Dommach, M. and Donat, A. and Dong, X. and Doynikov, N. and Dressel, M. and Duda, M. and Duda, P. and Eckoldt, H. and Ehsan, W. and Eidam, J. and Eints, F. and Engling, C. and Englisch, U. and Ermakov, A. and Escherich, K. and Eschke, J. and Saldin, E. and Faesing, M. and Fallou, A. and Felber, M. and Fenner, M. and Fernandes, B. and Fernández, J. M. and Feuker, S. and Filippakopoulos, K. and Floettmann, K. and Fogel, V. and Fontaine, M. and Francés, A. and Martin, I. Freijo and Freund, W. and Freyermuth, T. and Friedland, M. and Fröhlich, L. and Fusetti, M. and Fydrych, J. and Gallas, A. and García, O. and Garcia-Tabares, L. and Geloni, G. and Gerasimova, N. and Gerth, C. and Geßler, P. and Gharibyan, V. and Gloor, M. and Głowinkowski, J. and Goessel, A. and Golebiewski, Z. and Golubeva, N. and Grabowski, W. and Graeff, W. and Grebentsov, A. and Grecki, M. and Grevsmuehl, T. and Gross, M. and Grosse-Wortmann, U. and Grünert, J. and Grunewald, S. and Grzegory, P. and Feng, G. and Guler, H. and Gusev, G. and Gutierrez, J. L. and Hagge, L. and Hamberg, M. and Hanneken, R. and Harms, E. and Hartl, I. and Hauberg, A. and Hauf, S. and Hauschildt, J. and Hauser, J. and Havlicek, J. and Hedqvist, A. and Heidbrook, N. and Hellberg, F. and Henning, D. and Hensler, O. and Hermann, T. and Hidvégi, A. and Hierholzer, M. and Hintz, H. and Hoffmann, F. and Hoffmann, Markus and Hoffmann, Matthias and Holler, Y. and Hüning, M. and Ignatenko, A. and Ilchen, M. and Iluk, A. and Iversen, J. and Iversen, J. and Izquierdo, M. and Jachmann, L. and Jardon, N. and Jastrow, U. and Jensch, K. and Jensen, J. and Jeżabek, M. and Jidda, M. and Jin, H. and Johansson, N. and Jonas, R. and Kaabi, W. and Kaefer, D. and Kammering, R. and Kapitza, H. and Karabekyan, S. and Karstensen, S. and Kasprzak, K. and Katalev, V. and Keese, D. and Keil, B. and Kholopov, M. and Killenberger, M. and Kitaev, B. and Klimchenko, Y. and Klos, R. and Knebel, L. and Koch, A. and Koepke, M. and Köhler, S. and Köhler, W. and Kohlstrunk, N. and Konopkova, Z. and Konstantinov, A. and Kook, W. and Koprek, W. and Körfer, M. and Korth, O. and Kosarev, A. and Kosiński, K. and Kostin, D. and Kot, Y. and Kotarba, A. and Kozak, T. and Kozak, V. and Kramert, R. and Krasilnikov, M. and Krasnov, A. and Krause, B. and Kravchuk, L. and Krebs, O. and Kretschmer, R. and Kreutzkamp, J. and Kröplin, O. and Krzysik, K. and Kube, G. and Kuehn, H. and Kujala, N. and Kulikov, V. and Kuzminych, V. and La Civita, D. and Lacroix, M. and Lamb, T. and Lancetov, A. and Larsson, M. and Le Pinvidic, D. and Lederer, S. and Lensch, T. and Lenz, D. and Leuschner, A. and Levenhagen, F. and Li, Y. and Liebing, J. and Lilje, L. and Limberg, T. and Lipka, D. and List, B. and Liu, J. and Liu, S. and Lorbeer, B. and Lorkiewicz, J. and Lu, H. H. and Ludwig, F. and Machau, K. and Maciocha, W. and Madec, C. and Magueur, C. and Maiano, C. and Maksimova, I. and Malcher, K. and Maltezopoulos, T. and Mamoshkina, E. and Manschwetus, B. and Marcellini, F. and Marinkovic, G. and Martinez, T. and Martirosyan, H. and Maschmann, W. and Maslov, M. and Matheisen, A. and Mavric, U. and Meißner, J. and Meissner, K. and Messerschmidt, M. and Meyners, N. and Michalski, G. and Michelato, P. and Mildner, N. and Moe, M. and Moglia, F. and Mohr, C. and Mohr, S. and Möller, W. and Mommerz, M. and Monaco, L. and Montiel, C. and Moretti, M. and Morozov, I. and Morozov, P. and Mross, D. and Mueller, J. and Müller, C. and Müller, J. and Müller, K. and Munilla, J. and Münnich, A. and Muratov, V. and Napoly, O. and Näser, B. and Nefedov, N. and Neumann, Reinhard and Neumann, Rudolf and Ngada, N. and Noelle, D. and Obier, F. and Okunev, I. and Oliver, J. A. and Omet, M. and Oppelt, A. and Ottmar, A. and Oublaid, M. and Pagani, C. and Paparella, R. and Paramonov, V. and Peitzmann, C. and Penning, J. and Perus, A. and Peters, F. and Petersen, B. and Petrov, A. and Petrov, I. and Pfeiffer, S. and Pflüger, J. and Philipp, S. and Pienaud, Y. and Pierini, P. and Pivovarov, S. and Planas, M. and Pławski, E. and Pohl, M. and Polinski, J. and Popov, V. and Prat, S. and Prenting, J. and Priebe, G. and Pryschelski, H. and Przygoda, K. and Pyata, E. and Racky, B. and Rathjen, A. and Ratuschni, W. and Regnaud-Campderros, S. and Rehlich, K. and Reschke, D. and Robson, C. and Roever, J. and Roggli, M. and Rothenburg, J. and Rusiński, E. and Rybaniec, R. and Sahling, H. and Salmani, M. and Samoylova, L. and Sanzone, D. and Saretzki, F. and Sawlanski, O. and Schaffran, J. and Schlarb, H. and Schlösser, M. and Schlott, V. and Schmidt, C. and Schmidt-Foehre, F. and Schmitz, M. and Schmökel, M. and Schnautz, T. and Schneidmiller, E. and Scholz, M. and Schöneburg, B. and Schultze, J. and Schulz, C. and Schwarz, A. and Sekutowicz, J. and Sellmann, D. and Semenov, E. and Serkez, S. and Sertore, D. and Shehzad, N. and Shemarykin, P. and Shi, L. and Sienkiewicz, M. and Sikora, D. and Sikorski, M. and Silenzi, A. and Simon, C. and Singer, W. and Singer, X. and Sinn, H. and Sinram, K. and Skvorodnev, N. and Smirnow, P. and Sommer, T. and Sorokin, A. and Stadler, M. and Steckel, M. and Steffen, B. and Steinhau-Kühl, N. and Stephan, F. and Stodulski, M. and Stolper, M. and Sulimov, A. and Susen, R. and Świerblewski, J. and Sydlo, C. and Syresin, E. and Sytchev, V. and Szuba, J. and Tesch, N. and Thie, J. and Thiebault, A. and Tiedtke, K. and Tischhauser, D. and Tolkiehn, J. and Tomin, S. and Tonisch, F. and Toral, F. and Torbin, I. and Trapp, A. and Treyer, D. and Trowitzsch, G. and Trublet, T. and Tschentscher, T. and Ullrich, F. and Vannoni, M. and Varela, P. and Varghese, G. and Vashchenko, G. and Vasic, M. and Vazquez-Velez, C. and Verguet, A. and Vilcins-Czvitkovits, S. and Villanueva, R. and Visentin, B. and Viti, M. and Vogel, E. and Volobuev, E. and Wagner, R. and Walker, N. and Wamsat, T. and Weddig, H. and Weichert, G. and Weise, H. and Wenndorf, R. and Werner, M. and Wichmann, R. and Wiebers, C. and Wiencek, M. and Wilksen, T. and Will, I. and Winkelmann, L. and Winkowski, M. and Wittenburg, K. and Witzig, A. and Wlk, P. and Wohlenberg, T. and Wojciechowski, M. and Wolff-Fabris, F. and Wrochna, G. and Wrona, K. and Yakopov, M. and Yang, B. and Yang, F. and Yurkov, M. and Zagorodnov, I. and Zalden, P. and Zavadtsev, A. and Zavadtsev, D. and Zhirnov, A. and Zhukov, A. and Ziemann, V. and Zolotov, A. and Zolotukhina, N. and Zummack, F. and Zybin, D.},
	month = jun,
	year = {2020},
	pages = {391--397},
}

@article{barbatti_newton-x_2014,
	title = {Newton-{X}: a surface-hopping program for nonadiabatic molecular dynamics},
	volume = {4},
	copyright = {© 2013 John Wiley \& Sons, Ltd.},
	issn = {1759-0884},
	shorttitle = {Newton-{X}},
	url = {https://onlinelibrary.wiley.com/doi/abs/10.1002/wcms.1158},
	doi = {10.1002/wcms.1158},
	number = {1},
	urldate = {2025-02-03},
	journal = {WIREs Computational Molecular Science},
	author = {Barbatti, Mario and Ruckenbauer, Matthias and Plasser, Felix and Pittner, Jiri and Granucci, Giovanni and Persico, Maurizio and Lischka, Hans},
	year = {2014},
	note = {\_eprint: https://onlinelibrary.wiley.com/doi/pdf/10.1002/wcms.1158},
	pages = {26--33},
}

@article{oboyle_open_2011,
	title = {Open {Babel}: {An} open chemical toolbox},
	volume = {3},
	issn = {1758-2946},
	shorttitle = {Open {Babel}},
	url = {https://doi.org/10.1186/1758-2946-3-33},
	doi = {10.1186/1758-2946-3-33},
	number = {1},
	urldate = {2025-02-03},
	journal = {Journal of Cheminformatics},
	author = {O'Boyle, Noel M. and Banck, Michael and James, Craig A. and Morley, Chris and Vandermeersch, Tim and Hutchison, Geoffrey R.},
	month = oct,
	year = {2011},
	pages = {33},
}

@article{hanwell_avogadro_2012,
	title = {Avogadro: an advanced semantic chemical editor, visualization, and analysis platform},
	volume = {4},
	issn = {1758-2946},
	shorttitle = {Avogadro},
	url = {https://doi.org/10.1186/1758-2946-4-17},
	doi = {10.1186/1758-2946-4-17},
	number = {1},
	urldate = {2025-02-03},
	journal = {Journal of Cheminformatics},
	author = {Hanwell, Marcus D. and Curtis, Donald E. and Lonie, David C. and Vandermeersch, Tim and Zurek, Eva and Hutchison, Geoffrey R.},
	month = aug,
	year = {2012},
	pages = {17},
}

@article{yuan_coulomb_2024,
	title = {Coulomb {Explosion} {Imaging} of {Complex} {Molecules} {Using} {Highly} {Charged} {Ions}},
	volume = {133},
	url = {https://link.aps.org/doi/10.1103/PhysRevLett.133.193002},
	doi = {10.1103/PhysRevLett.133.193002},
	number = {19},
	urldate = {2024-11-27},
	journal = {Physical Review Letters},
	author = {Yuan, Hang and Gao, Yue and Yang, Bo and Gu, Shaofei and Lin, Hong and Guo, Dalong and Liu, Junliang and Zhang, Shaofeng and Ma, Xinwen and Xu, Shenyue},
	month = nov,
	year = {2024},
	note = {Publisher: American Physical Society},
	pages = {193002},
}

@article{bhattacharyya_strong-field-induced_2022,
	title = {Strong-{Field}-{Induced} {Coulomb} {Explosion} {Imaging} of {Tribromomethane}},
	volume = {13},
	url = {https://doi.org/10.1021/acs.jpclett.2c01007},
	doi = {10.1021/acs.jpclett.2c01007},
	number = {25},
	urldate = {2024-07-23},
	journal = {The Journal of Physical Chemistry Letters},
	author = {Bhattacharyya, Surjendu and Borne, Kurtis and Ziaee, Farzaneh and Pathak, Shashank and Wang, Enliang and Venkatachalam, Anbu Selvam and Li, Xiang and Marshall, Nathan and Carnes, Kevin D. and Fehrenbach, Charles W. and Severt, Travis and Ben-Itzhak, Itzik and Rudenko, Artem and Rolles, Daniel},
	month = jun,
	year = {2022},
	pages = {5845--5853},
}

@article{centurion_ultrafast_2022,
	title = {Ultrafast {Imaging} of {Molecules} with {Electron} {Diffraction}},
	volume = {73},
	issn = {0066-426X, 1545-1593},
	url = {https://www.annualreviews.org/content/journals/10.1146/annurev-physchem-082720-010539},
	doi = {10.1146/annurev-physchem-082720-010539},
	language = {en},
	number = {Volume 73, 2022},
	urldate = {2024-07-15},
	journal = {Annual Review of Physical Chemistry},
	author = {Centurion, Martin and Wolf, Thomas J. A. and Yang, Jie},
	month = apr,
	year = {2022},
	pages = {21--42},
}

@article{moreno_carrascosa_ab_2019,
	title = {Ab {Initio} {Calculation} of {Total} {X}-ray {Scattering} from {Molecules}},
	volume = {15},
	issn = {1549-9618},
	url = {https://doi.org/10.1021/acs.jctc.9b00056},
	doi = {10.1021/acs.jctc.9b00056},
	number = {5},
	urldate = {2024-07-15},
	journal = {Journal of Chemical Theory and Computation},
	author = {Moreno Carrascosa, Andrés and Yong, Haiwang and Crittenden, Deborah L. and Weber, Peter M. and Kirrander, Adam},
	month = may,
	year = {2019},
	pages = {2836--2846},
}

@article{hubbell_atomic_1975,
	title = {Atomic form factors, incoherent scattering functions, and photon scattering cross sections},
	volume = {4},
	issn = {0047-2689},
	url = {https://doi.org/10.1063/1.555523},
	doi = {10.1063/1.555523},
	number = {3},
	urldate = {2024-07-15},
	journal = {Journal of Physical and Chemical Reference Data},
	author = {Hubbell, J. H. and Veigele, Wm. J. and Briggs, E. A. and Brown, R. T. and Cromer, D. T. and Howerton, R. J.},
	month = jul,
	year = {1975},
	pages = {471--538},
}

@article{ruddock_deep_2019,
	title = {A deep {UV} trigger for ground-state ring-opening dynamics of 1,3-cyclohexadiene},
	volume = {5},
	issn = {2375-2548},
	url = {https://www.science.org/doi/10.1126/sciadv.aax6625},
	doi = {10.1126/sciadv.aax6625},
	language = {en},
	number = {9},
	urldate = {2024-07-15},
	journal = {Science Advances},
	author = {Ruddock, Jennifer M. and Yong, Haiwang and Stankus, Brian and Du, Wenpeng and Goff, Nathan and Chang, Yu and Odate, Asami and Carrascosa, Andrés Moreno and Bellshaw, Darren and Zotev, Nikola and Liang, Mengning and Carbajo, Sergio and Koglin, Jason and Robinson, Joseph S. and Boutet, Sébastien and Kirrander, Adam and Minitti, Michael P. and Weber, Peter M.},
	month = sep,
	year = {2019},
	pages = {eaax6625},
}

@article{zhou_coulomb_2020,
	title = {Coulomb explosion imaging for gas-phase molecular structure determination: {An} \textit{ab initio} trajectory simulation study},
	volume = {153},
	issn = {0021-9606, 1089-7690},
	shorttitle = {Coulomb explosion imaging for gas-phase molecular structure determination},
	url = {https://pubs.aip.org/jcp/article/153/18/184201/1062437/Coulomb-explosion-imaging-for-gas-phase-molecular},
	doi = {10.1063/5.0024833},
	language = {en},
	number = {18},
	urldate = {2024-07-11},
	journal = {The Journal of Chemical Physics},
	author = {Zhou, Weiwei and Ge, Lingfeng and Cooper, Graham A. and Crane, Stuart W. and Evans, Michael H. and Ashfold, Michael N. R. and Vallance, Claire},
	month = nov,
	year = {2020},
	pages = {184201},
}

@article{wang_time-resolved_nodate,
	title = {Time-{Resolved} {Coulomb} {Explosion} {Imaging} {Unveils} {Ultrafast} {Ring} {Opening} of {Furan}},
	url = {https://arxiv.org/abs/2311.05099},
	language = {en},
	author = {Wang, Enliang and Bhattacharyya, Surjendu and Chen, Keyu and Borne, Kurtis and Pathak, Shashank and Lam, Huynh Van Sa and Venkatachalam, Anbu Selvam and Chen, Xiangjun and Boll, Rebecca and Jahnke, Till and Rudenko, Artem and Rolles, Daniel},
    year = {2023},
    journal={https://arxiv.org/abs/2311.05099},
}

@article{herwig_metastable_2013,
	title = {Metastable states of {D2}- observed by foil-induced {Coulomb} explosion imaging},
	volume = {87},
	url = {https://link.aps.org/doi/10.1103/PhysRevA.87.062513},
	doi = {10.1103/PhysRevA.87.062513},
	number = {6},
	urldate = {2024-04-21},
	journal = {Physical Review A},
	author = {Herwig, P. and Schwalm, D. and Čížek, M. and Golser, R. and Grieser, M. and Heber, O. and Repnow, R. and Wolf, A. and Kreckel, H.},
	month = jun,
	year = {2013},
	pages = {062513},
}

@article{wang_ultrafast_2020,
	title = {Ultrafast {Proton} {Transfer} {Dynamics} on the {Repulsive} {Potential} of the {Ethanol} {Dication}: {Roaming}-{Mediated} {Isomerization} versus {Coulomb} {Explosion}},
	volume = {124},
	issn = {1089-5639},
	shorttitle = {Ultrafast {Proton} {Transfer} {Dynamics} on the {Repulsive} {Potential} of the {Ethanol} {Dication}},
	url = {https://doi.org/10.1021/acs.jpca.0c02074},
	doi = {10.1021/acs.jpca.0c02074},
	number = {14},
	urldate = {2024-04-21},
	journal = {The Journal of Physical Chemistry A},
	author = {Wang, Enliang and Shan, Xu and Chen, Lei and Pfeifer, Thomas and Chen, Xiangjun and Ren, Xueguang and Dorn, Alexander},
	month = apr,
	year = {2020},
	pages = {2785--2791},
}

@article{lam_differentiating_2024,
	title = {Differentiating {Three}-{Dimensional} {Molecular} {Structures} {Using} {Laser}-{Induced} {Coulomb} {Explosion} {Imaging}},
	volume = {132},
	url = {https://link.aps.org/doi/10.1103/PhysRevLett.132.123201},
	doi = {10.1103/PhysRevLett.132.123201},
	number = {12},
	urldate = {2024-04-12},
	journal = {Physical Review Letters},
	author = {Lam, Huynh Van Sa and Venkatachalam, Anbu Selvam and Bhattacharyya, Surjendu and Chen, Keyu and Borne, Kurtis and Wang, Enliang and Boll, Rebecca and Jahnke, Till and Kumarappan, Vinod and Rudenko, Artem and Rolles, Daniel},
	month = mar,
	year = {2024},
	pages = {123201},
}

@article{wolf_photochemical_2019,
	title = {The photochemical ring-opening of 1,3-cyclohexadiene imaged by ultrafast electron diffraction},
	volume = {11},
	copyright = {2019 This is a U.S. government work and not under copyright protection in the U.S.; foreign copyright protection may apply},
	issn = {1755-4349},
	url = {https://www.nature.com/articles/s41557-019-0252-7},
	doi = {10.1038/s41557-019-0252-7},
	language = {en},
	number = {6},
	urldate = {2023-06-10},
	journal = {Nature Chemistry},
	author = {Wolf, T. J. A. and Sanchez, D. M. and Yang, J. and Parrish, R. M. and Nunes, J. P. F. and Centurion, M. and Coffee, R. and Cryan, J. P. and Gühr, M. and Hegazy, K. and Kirrander, A. and Li, R. K. and Ruddock, J. and Shen, X. and Vecchione, T. and Weathersby, S. P. and Weber, P. M. and Wilkin, K. and Yong, H. and Zheng, Q. and Wang, X. J. and Minitti, M. P. and Martínez, T. J.},
	month = jun,
	year = {2019},
	pages = {504--509},
}

@article{endo_capturing_2020,
	title = {Capturing roaming molecular fragments in real time},
	volume = {370},
	url = {https://www.science.org/doi/10.1126/science.abc2960},
	doi = {10.1126/science.abc2960},
	number = {6520},
	urldate = {2024-04-20},
	journal = {Science},
	author = {Endo, Tomoyuki and Neville, Simon P. and Wanie, Vincent and Beaulieu, Samuel and Qu, Chen and Deschamps, Jude and Lassonde, Philippe and Schmidt, Bruno E. and Fujise, Hikaru and Fushitani, Mizuho and Hishikawa, Akiyoshi and Houston, Paul L. and Bowman, Joel M. and Schuurman, Michael S. and Légaré, François and Ibrahim, Heide},
	month = nov,
	year = {2020},
	pages = {1072--1077},
}

@article{pickering_communication_2016,
	title = {Communication: {Three}-fold covariance imaging of laser-induced {Coulomb} explosions},
	volume = {144},
	issn = {0021-9606},
	shorttitle = {Communication},
	url = {https://doi.org/10.1063/1.4947551},
	doi = {10.1063/1.4947551},
	number = {16},
	urldate = {2024-04-21},
	journal = {The Journal of Chemical Physics},
	author = {Pickering, James D. and Amini, Kasra and Brouard, Mark and Burt, Michael and Bush, Ian J. and Christensen, Lauge and Lauer, Alexandra and Nielsen, Jens H. and Slater, Craig S. and Stapelfeldt, Henrik},
	month = apr,
	year = {2016},
	pages = {161105},
}

@article{schmidt_spatial_2012,
	title = {Spatial {Imaging} of the {H} 2 + {Vibrational} {Wave} {Function} at the {Quantum} {Limit}},
	volume = {108},
	copyright = {http://link.aps.org/licenses/aps-default-license},
	issn = {0031-9007, 1079-7114},
	url = {https://link.aps.org/doi/10.1103/PhysRevLett.108.073202},
	doi = {10.1103/PhysRevLett.108.073202},
	language = {en},
	number = {7},
	urldate = {2024-04-21},
	journal = {Physical Review Letters},
	author = {Schmidt, L. Ph. H. and Jahnke, T. and Czasch, A. and Schöffler, M. and Schmidt-Böcking, H. and Dörner, R.},
	month = feb,
	year = {2012},
	pages = {073202},
}

@article{neumann_fragmentation_2010,
	title = {Fragmentation {Dynamics} of {CO} 2 3 + {Investigated} by {Multiple} {Electron} {Capture} in {Collisions} with {Slow} {Highly} {Charged} {Ions}},
	volume = {104},
	copyright = {http://link.aps.org/licenses/aps-default-license},
	issn = {0031-9007, 1079-7114},
	url = {https://link.aps.org/doi/10.1103/PhysRevLett.104.103201},
	doi = {10.1103/PhysRevLett.104.103201},
	language = {en},
	number = {10},
	urldate = {2024-04-21},
	journal = {Physical Review Letters},
	author = {Neumann, N. and Hant, D. and Schmidt, L. Ph. H. and Titze, J. and Jahnke, T. and Czasch, A. and Schöffler, M. S. and Kreidi, K. and Jagutzki, O. and Schmidt-Böcking, H. and Dörner, R.},
	month = mar,
	year = {2010},
	pages = {103201},
}

@article{gemmell_determining_1980,
	title = {Determining the stereochemical structures of molecular ions by "{Coulomb}-explosion" techniques with fast ({MeV}) molecular ion beams},
	volume = {80},
	issn = {0009-2665, 1520-6890},
	url = {https://pubs.acs.org/doi/abs/10.1021/cr60326a002},
	doi = {10.1021/cr60326a002},
	language = {en},
	number = {4},
	urldate = {2024-04-21},
	journal = {Chemical Reviews},
	author = {Gemmell, Donald S.},
	month = aug,
	year = {1980},
	pages = {301--311},
}

@article{kastirke_double_2020,
	title = {Double {Core}-{Hole} {Generation} in {O} 2 {Molecules} {Using} an {X}-{Ray} {Free}-{Electron} {Laser}: {Molecular}-{Frame} {Photoelectron} {Angular} {Distributions}},
	volume = {125},
	issn = {0031-9007, 1079-7114},
	shorttitle = {Double {Core}-{Hole} {Generation} in {O} 2 {Molecules} {Using} an {X}-{Ray} {Free}-{Electron} {Laser}},
	url = {https://link.aps.org/doi/10.1103/PhysRevLett.125.163201},
	doi = {10.1103/PhysRevLett.125.163201},
	language = {en},
	number = {16},
	urldate = {2024-04-10},
	journal = {Physical Review Letters},
	author = {Kastirke, Gregor and Schöffler, Markus S. and Weller, Miriam and Rist, Jonas and Boll, Rebecca and Anders, Nils and Baumann, Thomas M. and Eckart, Sebastian and Erk, Benjamin and De Fanis, Alberto and Fehre, Kilian and Gatton, Averell and Grundmann, Sven and Grychtol, Patrik and Hartung, Alexander and Hofmann, Max and Ilchen, Markus and Janke, Christian and Kircher, Max and Kunitski, Maksim and Li, Xiang and Mazza, Tommaso and Melzer, Niklas and Montano, Jacobo and Music, Valerija and Nalin, Giammarco and Ovcharenko, Yevheniy and Pier, Andreas and Rennhack, Nils and Rivas, Daniel E. and Dörner, Reinhard and Rolles, Daniel and Rudenko, Artem and Schmidt, Philipp and Siebert, Juliane and Strenger, Nico and Trabert, Daniel and Vela-Perez, Isabel and Wagner, Rene and Weber, Thorsten and Williams, Joshua B. and Ziolkowski, Pawel and Schmidt, Lothar Ph. H. and Czasch, Achim and Ueda, Kiyoshi and Trinter, Florian and Meyer, Michael and Demekhin, Philipp V. and Jahnke, Till},
	month = oct,
	year = {2020},
	pages = {163201},
}

@article{kastirke_photoelectron_2020,
	title = {Photoelectron {Diffraction} {Imaging} of a {Molecular} {Breakup} {Using} an {X}-{Ray} {Free}-{Electron} {Laser}},
	volume = {10},
	issn = {2160-3308},
	url = {https://link.aps.org/doi/10.1103/PhysRevX.10.021052},
	doi = {10.1103/PhysRevX.10.021052},
	language = {en},
	number = {2},
	urldate = {2024-04-10},
	journal = {Physical Review X},
	author = {Kastirke, Gregor and Schöffler, Markus S. and Weller, Miriam and Rist, Jonas and Boll, Rebecca and Anders, Nils and Baumann, Thomas M. and Eckart, Sebastian and Erk, Benjamin and De Fanis, Alberto and Fehre, Kilian and Gatton, Averell and Grundmann, Sven and Grychtol, Patrik and Hartung, Alexander and Hofmann, Max and Ilchen, Markus and Janke, Christian and Kircher, Max and Kunitski, Maksim and Li, Xiang and Mazza, Tommaso and Melzer, Niklas and Montano, Jacobo and Music, Valerija and Nalin, Giammarco and Ovcharenko, Yevheniy and Pier, Andreas and Rennhack, Nils and Rivas, Daniel E. and Dörner, Reinhard and Rolles, Daniel and Rudenko, Artem and Schmidt, Philipp and Siebert, Juliane and Strenger, Nico and Trabert, Daniel and Vela-Perez, Isabel and Wagner, Rene and Weber, Thorsten and Williams, Joshua B. and Ziolkowski, Pawel and Schmidt, Lothar Ph. H. and Czasch, Achim and Trinter, Florian and Meyer, Michael and Ueda, Kiyoshi and Demekhin, Philipp V. and Jahnke, Till},
	month = jun,
	year = {2020},
	pages = {021052},
}

@article{restrepo_alkyl_2007,
	title = {Alkyl chains with {CN} and {CCH} substituents prefer gauche conformations},
	volume = {833},
	copyright = {https://www.elsevier.com/tdm/userlicense/1.0/},
	issn = {00222860},
	url = {https://linkinghub.elsevier.com/retrieve/pii/S0022286006007502},
	doi = {10.1016/j.molstruc.2006.09.020},
	language = {en},
	number = {1-3},
	urldate = {2024-04-05},
	journal = {Journal of Molecular Structure},
	author = {Restrepo, Albeiro A. and Bohn, Robert K.},
	month = may,
	year = {2007},
	pages = {189--196},
}

@article{cheng_multiparticle_2023,
	title = {Multiparticle {Cumulant} {Mapping} for {Coulomb} {Explosion} {Imaging}},
	volume = {130},
	issn = {0031-9007, 1079-7114},
	url = {https://link.aps.org/doi/10.1103/PhysRevLett.130.093001},
	doi = {10.1103/PhysRevLett.130.093001},
	language = {en},
	number = {9},
	urldate = {2023-03-06},
	journal = {Physical Review Letters},
	author = {Cheng, Chuan and Frasinski, Leszek J. and Moğol, Gönenç and Allum, Felix and Howard, Andrew J. and Rolles, Daniel and Bucksbaum, Philip H. and Brouard, Mark and Forbes, Ruaridh and Weinacht, Thomas},
	month = mar,
	year = {2023},
	pages = {093001},
}

@article{schouder_laser-induced_2022,
	title = {Laser-{Induced} {Coulomb} {Explosion} {Imaging} of {Aligned} {Molecules} and {Molecular} {Dimers}},
	volume = {73},
	url = {https://doi.org/10.1146/annurev-physchem-090419-053627},
	doi = {10.1146/annurev-physchem-090419-053627},
	number = {1},
	urldate = {2023-07-18},
	journal = {Annual Review of Physical Chemistry},
	author = {Schouder, Constant A. and Chatterley, Adam S. and Pickering, James D. and Stapelfeldt, Henrik},
	year = {2022},
	pmid = {35081323},
	note = {\_eprint: https://doi.org/10.1146/annurev-physchem-090419-053627},
	pages = {323--347},
}

@article{christensen_dynamic_2014,
	title = {Dynamic {Stark} {Control} of {Torsional} {Motion} by a {Pair} of {Laser} {Pulses}},
	volume = {113},
	issn = {0031-9007, 1079-7114},
	url = {https://link.aps.org/doi/10.1103/PhysRevLett.113.073005},
	doi = {10.1103/PhysRevLett.113.073005},
	language = {en},
	number = {7},
	urldate = {2022-12-07},
	journal = {Physical Review Letters},
	author = {Christensen, Lauge and Nielsen, Jens H. and Brandt, Christian B. and Madsen, Christian B. and Madsen, Lars Bojer and Slater, Craig S. and Lauer, Alexandra and Brouard, Mark and Johansson, Mikael P. and Shepperson, Benjamin and Stapelfeldt, Henrik},
	month = aug,
	year = {2014},
	pages = {073005},
}

@article{li_coulomb_2022,
	title = {Coulomb explosion imaging of small polyatomic molecules with ultrashort x-ray pulses},
	volume = {4},
	issn = {2643-1564},
	url = {https://link.aps.org/doi/10.1103/PhysRevResearch.4.013029},
	doi = {10.1103/PhysRevResearch.4.013029},
	language = {en},
	number = {1},
	urldate = {2022-05-02},
	journal = {Physical Review Research},
	author = {Li, X. and Rudenko, A. and Schöffler, M. S. and Anders, N. and Baumann, Th. M. and Eckart, S. and Erk, B. and De Fanis, A. and Fehre, K. and Dörner, R. and Foucar, L. and Grundmann, S. and Grychtol, P. and Hartung, A. and Hofmann, M. and Ilchen, M. and Janke, Ch. and Kastirke, G. and Kircher, M. and Kubicek, K. and Kunitski, M. and Mazza, T. and Meister, S. and Melzer, N. and Montano, J. and Music, V. and Nalin, G. and Ovcharenko, Y. and Passow, Ch. and Pier, A. and Rennhack, N. and Rist, J. and Rivas, D. E. and Schlichting, I. and Schmidt, L. Ph. H. and Schmidt, Ph. and Siebert, J. and Strenger, N. and Trabert, D. and Trinter, F. and Vela-Perez, I. and Wagner, R. and Walter, P. and Weller, M. and Ziolkowski, P. and Czasch, A. and Rolles, D. and Meyer, M. and Jahnke, T. and Boll, R.},
	month = jan,
	year = {2022},
	pages = {013029},
}

@article{boll_x-ray_2022,
	title = {X-ray multiphoton-induced {Coulomb} explosion images complex single molecules},
	volume = {18},
	issn = {1745-2473, 1745-2481},
	url = {https://www.nature.com/articles/s41567-022-01507-0},
	doi = {10.1038/s41567-022-01507-0},
	language = {en},
	number = {4},
	urldate = {2022-05-02},
	journal = {Nature Physics},
	author = {Boll, Rebecca and Schäfer, Julia M. and Richard, Benoît and Fehre, Kilian and Kastirke, Gregor and Jurek, Zoltan and Schöffler, Markus S. and Abdullah, Malik M. and Anders, Nils and Baumann, Thomas M. and Eckart, Sebastian and Erk, Benjamin and De Fanis, Alberto and Dörner, Reinhard and Grundmann, Sven and Grychtol, Patrik and Hartung, Alexander and Hofmann, Max and Ilchen, Markus and Inhester, Ludger and Janke, Christian and Jin, Rui and Kircher, Max and Kubicek, Katharina and Kunitski, Maksim and Li, Xiang and Mazza, Tommaso and Meister, Severin and Melzer, Niklas and Montano, Jacobo and Music, Valerija and Nalin, Giammarco and Ovcharenko, Yevheniy and Passow, Christopher and Pier, Andreas and Rennhack, Nils and Rist, Jonas and Rivas, Daniel E. and Rolles, Daniel and Schlichting, Ilme and Schmidt, Lothar Ph. H. and Schmidt, Philipp and Siebert, Juliane and Strenger, Nico and Trabert, Daniel and Trinter, Florian and Vela-Perez, Isabel and Wagner, Rene and Walter, Peter and Weller, Miriam and Ziolkowski, Pawel and Son, Sang-Kil and Rudenko, Artem and Meyer, Michael and Santra, Robin and Jahnke, Till},
	month = apr,
	year = {2022},
	pages = {423--428},
}

@article{pathak_differentiating_2020,
	title = {Differentiating and {Quantifying} {Gas}-{Phase} {Conformational} {Isomers} {Using} {Coulomb} {Explosion} {Imaging}},
	volume = {11},
	issn = {1948-7185, 1948-7185},
	url = {https://pubs.acs.org/doi/10.1021/acs.jpclett.0c02959},
	doi = {10.1021/acs.jpclett.0c02959},
	language = {en},
	number = {23},
	urldate = {2021-05-25},
	journal = {The Journal of Physical Chemistry Letters},
	author = {Pathak, Shashank and Obaid, Razib and Bhattacharyya, Surjendu and Bürger, Johannes and Li, Xiang and Tross, Jan and Severt, Travis and Davis, Brandin and Bilodeau, René C. and Trallero-Herrero, Carlos A. and Rudenko, Artem and Berrah, Nora and Rolles, Daniel},
	month = dec,
	year = {2020},
	pages = {10205--10211},
}

@article{mcdonnell_ultrafast_2020,
	title = {Ultrafast {Laser}-{Induced} {Isomerization} {Dynamics} in {Acetonitrile}},
	issn = {1948-7185, 1948-7185},
	url = {https://pubs.acs.org/doi/10.1021/acs.jpclett.0c01344},
	doi = {10.1021/acs.jpclett.0c01344},
	language = {en},
	urldate = {2020-08-09},
	journal = {The Journal of Physical Chemistry Letters},
	author = {McDonnell, Matteo and LaForge, Aaron C. and Reino-González, Juan and Disla, Martin and Kling, Nora G. and Mishra, Debadarshini and Obaid, Razib and Sundberg, Margaret and Svoboda, Vít and Díaz-Tendero, Sergio and Martín, Fernando and Berrah, Nora},
	month = aug,
	year = {2020},
    volume = {11},
	pages = {6724--6729},
}

@article{hishikawa_visualizing_2007,
	title = {Visualizing {Recurrently} {Migrating} {Hydrogen} in {Acetylene} {Dication} by {Intense} {Ultrashort} {Laser} {Pulses}},
	volume = {99},
	issn = {0031-9007, 1079-7114},
	url = {https://link.aps.org/doi/10.1103/PhysRevLett.99.258302},
	doi = {10.1103/PhysRevLett.99.258302},
	language = {en},
	number = {25},
	urldate = {2020-05-14},
	journal = {Physical Review Letters},
	author = {Hishikawa, Akiyoshi and Matsuda, Akitaka and Fushitani, Mizuho and Takahashi, Eiji J.},
	month = dec,
	year = {2007},
	pages = {258302},
}

@article{pitzer_direct_2013,
	title = {Direct {Determination} of {Absolute} {Molecular} {Stereochemistry} in {Gas} {Phase} by {Coulomb} {Explosion} {Imaging}},
	volume = {341},
	issn = {0036-8075, 1095-9203},
	url = {https://www.sciencemag.org/lookup/doi/10.1126/science.1240362},
	doi = {10.1126/science.1240362},
	language = {en},
	number = {6150},
	urldate = {2020-05-04},
	journal = {Science},
	author = {Pitzer, M. and Kunitski, M. and Johnson, A. S. and Jahnke, T. and Sann, H. and Sturm, F. and Schmidt, L. P. H. and Schmidt-Bocking, H. and Dorner, R. and Stohner, J. and Kiedrowski, J. and Reggelin, M. and Marquardt, S. and Schiesser, A. and Berger, R. and Schoffler, M. S.},
	month = sep,
	year = {2013},
	pages = {1096--1100},
}

@article{pitzer_absolute_2016,
	title = {Absolute {Configuration} from {Different} {Multifragmentation} {Pathways} in {Light}-{Induced} {Coulomb} {Explosion} {Imaging}},
	volume = {17},
	issn = {14394235},
	url = {http://doi.wiley.com/10.1002/cphc.201501118},
	doi = {10.1002/cphc.201501118},
	language = {en},
	number = {16},
	urldate = {2020-05-03},
	journal = {ChemPhysChem},
	author = {Pitzer, Martin and Kastirke, Gregor and Kunitski, Maksim and Jahnke, Till and Bauer, Tobias and Goihl, Christoph and Trinter, Florian and Schober, Carl and Henrichs, Kevin and Becht, Jasper and Zeller, Stefan and Gassert, Helena and Waitz, Markus and Kuhlins, Andreas and Sann, Hendrik and Sturm, Felix and Wiegandt, Florian and Wallauer, Robert and Schmidt, Lothar Ph. H. and Johnson, Allan S. and Mazenauer, Manuel and Spenger, Benjamin and Marquardt, Sabrina and Marquardt, Sebastian and Schmidt-Böcking, Horst and Stohner, Jürgen and Dörner, Reinhard and Schöffler, Markus and Berger, Robert},
	month = aug,
	year = {2016},
	pages = {2465--2472},
}

@article{eland_dynamics_1987,
	title = {The dynamics of three-body dissociations of dications studied by the triple coincidence technique {PEPIPICO}},
	volume = {61},
	issn = {0026-8976, 1362-3028},
	url = {http://www.tandfonline.com/doi/abs/10.1080/00268978700101421},
	doi = {10.1080/00268978700101421},
	language = {en},
	number = {3},
	urldate = {2020-03-12},
	journal = {Molecular Physics},
	author = {Eland, J.H.D.},
	month = jun,
	year = {1987},
	pages = {725--745},
}

@article{amini_photodissociation_2018,
	title = {Photodissociation of aligned {CH} $_{\textrm{3}}$ {I} and {C} $_{\textrm{6}}$ {H} $_{\textrm{3}}$ {F} $_{\textrm{2}}$ {I} molecules probed with time-resolved {Coulomb} explosion imaging by site-selective extreme ultraviolet ionization},
	volume = {5},
	issn = {2329-7778},
	url = {http://aca.scitation.org/doi/10.1063/1.4998648},
	doi = {10.1063/1.4998648},
	language = {en},
	number = {1},
	urldate = {2020-03-01},
	journal = {Structural Dynamics},
	author = {Amini, Kasra and Savelyev, Evgeny and Brauße, Felix and Berrah, Nora and Bomme, Cédric and Brouard, Mark and Burt, Michael and Christensen, Lauge and Düsterer, Stefan and Erk, Benjamin and Höppner, Hauke and Kierspel, Thomas and Krecinic, Faruk and Lauer, Alexandra and Lee, Jason W. L. and Müller, Maria and Müller, Erland and Mullins, Terence and Redlin, Harald and Schirmel, Nora and Thøgersen, Jan and Techert, Simone and Toleikis, Sven and Treusch, Rolf and Trippel, Sebastian and Ulmer, Anatoli and Vallance, Claire and Wiese, Joss and Johnsson, Per and Küpper, Jochen and Rudenko, Artem and Rouzée, Arnaud and Stapelfeldt, Henrik and Rolles, Daniel and Boll, Rebecca},
	month = jan,
	year = {2018},
	pages = {014301},
}

@article{allum_coulomb_2018,
	title = {Coulomb explosion imaging of {CH} $_{\textrm{3}}$ {I} and {CH} $_{\textrm{2}}$ {ClI} photodissociation dynamics},
	volume = {149},
	issn = {0021-9606, 1089-7690},
	url = {http://aip.scitation.org/doi/10.1063/1.5041381},
	doi = {10.1063/1.5041381},
	language = {en},
	number = {20},
	urldate = {2019-11-06},
	journal = {The Journal of Chemical Physics},
	author = {Allum, Felix and Burt, Michael and Amini, Kasra and Boll, Rebecca and Köckert, Hansjochen and Olshin, Pavel K. and Bari, Sadia and Bomme, Cédric and Brauße, Felix and Cunha de Miranda, Barbara and Düsterer, Stefan and Erk, Benjamin and Géléoc, Marie and Geneaux, Romain and Gentleman, Alexander S. and Goldsztejn, Gildas and Guillemin, Renaud and Holland, David M. P. and Ismail, Iyas and Johnsson, Per and Journel, Loïc and Küpper, Jochen and Lahl, Jan and Lee, Jason W. L. and Maclot, Sylvain and Mackenzie, Stuart R. and Manschwetus, Bastian and Mereshchenko, Andrey S. and Mason, Robert and Palaudoux, Jérôme and Piancastelli, Maria Novella and Penent, Francis and Rompotis, Dimitrios and Rouzée, Arnaud and Ruchon, Thierry and Rudenko, Artem and Savelyev, Evgeny and Simon, Marc and Schirmel, Nora and Stapelfeldt, Henrik and Techert, Simone and Travnikova, Oksana and Trippel, Sebastian and Underwood, Jonathan G. and Vallance, Claire and Wiese, Joss and Ziaee, Farzaneh and Brouard, Mark and Marchenko, Tatiana and Rolles, Daniel},
	month = nov,
	year = {2018},
	pages = {204313},
}

@article{vager_coulomb_1989,
	title = {Coulomb {Explosion} {Imaging} of {Small} {Molecules}},
	volume = {244},
	issn = {0036-8075},
	url = {https://www.jstor.org/stable/1703084},
	number = {4903},
	urldate = {2019-03-16},
	journal = {Science},
	author = {Vager, Z. and Naaman, R. and Kanter, E. P.},
	year = {1989},
	pages = {426--431},
}

@article{ablikim_identification_2016,
	title = {Identification of absolute geometries of \textit{cis} and \textit{trans} molecular isomers by {Coulomb} {Explosion} {Imaging}},
	volume = {6},
	copyright = {2016 Nature Publishing Group},
	issn = {2045-2322},
	url = {https://www.nature.com/articles/srep38202},
	doi = {10.1038/srep38202},
	language = {en},
	urldate = {2019-03-16},
	journal = {Scientific Reports},
	author = {Ablikim, Utuq and Bomme, Cédric and Xiong, Hui and Savelyev, Evgeny and Obaid, Razib and Kaderiya, Balram and Augustin, Sven and Schnorr, Kirsten and Dumitriu, Ileana and Osipov, Timur and Bilodeau, René and Kilcoyne, David and Kumarappan, Vinod and Rudenko, Artem and Berrah, Nora and Rolles, Daniel},
	month = dec,
	year = {2016},
	pages = {38202},
}

@article{minitti_imaging_2015,
    author = {Minitti, M.P. and Budarz, J.M. and Kirrander, A. and Robinson, J.S. and Ratner, D. and Lane, T.J. and Zhu, D. and Glownia, J.M. and Kozina, M. and Lemke, H.T. and Sikorski, M. and Feng, Y. and Nelson, S. and Saita, K. and Stankus, B. and Northey, T. and Hastings, J.B. and Weber, P.M.},
    title = {Imaging Molecular Motion: Femtosecond X-Ray Scattering of an Electrocyclic Chemical Reaction},
    journal = {Physical Review Letters},
    year = {2015},
    volume = {114},
    number = {25},
    pages = {255501},
    month = jun,
    doi = {10.1103/PhysRevLett.114.255501},
    url = {https://link.aps.org/doi/10.1103/PhysRevLett.114.255501},
    language = {en}
}

@article{venkatachalam2025exploiting,
  title={Exploiting correlations in multi-coincidence Coulomb explosion patterns for differentiating molecular structures using machine learning},
  author={Venkatachalam, A. S. and Greenman, L. and Stallbaumer, J. and Rudenko, A. and Rolles, D. and Lam, H. V. S.},
  journal={Nature Communications},
  volume       = {16},
  pages        = {11366}, 
  year={2025},
}

@article{li2025imaging,
  title={Imaging a light-induced molecular elimination reaction with an {X}-ray free-electron laser},
  author={Li, Xiang and Boll, Rebecca and Vindel-Zandbergen, Patricia and Gonz{\'a}lez-V{\'a}zquez, Jes{\'u}s and Rivas, Daniel E and Bhattacharyya, Surjendu and Borne, Kurtis and Chen, Keyu and De Fanis, Alberto and Erk, Benjamin and others},
  journal={Nature Communications},
  volume={16},
  number={1},
  pages={7006},
  year={2025},
  publisher={Nature Publishing Group UK London}
}

@article{Lam_SO2, 
    title={Simultaneous imaging of vibrational, rotational, and electronic wave-packet dynamics in a triatomic molecule}, volume={111}, DOI={10.1103/1yd8-786g}, number={6}, journal={Physical Review A}, publisher={American Physical Society}, author={Lam, Huynh Van Sa and Hoang, Van-Hung and Venkatachalam, Anbu Selvam and Bhattacharyya, Surjendu and Chen, Keyu and Jacob, Sina and Kudagama, Sanduni and Nguyen, Tu Thanh and Rolles, Daniel and Thumm, Uwe and Rudenko, Artem and Kumarappan, Vinod}, year={2025}, month=june, pages={L061101} }

@article{li_AI_submitted,
  title={Generative Modeling Enables Molecular Structure Retrieval from Coulomb Explosion Imaging},
  author={Li, Xiang and Jahnke, Till and Boll, Rebecca and Han, Jiaqi and Xu, Minkai and Meyer, Michael and Piancastelli, Maria-Novella and Rolles, Daniel and Rudenko, Artem and Trinter, Florian and Wolf, Thomas J. A. and Thayer, Jana B. and Cryan, James P. and Ermont, Stefan and Ho, Phay J.},
  journal={https://www.arxiv.org/abs/2511.00179},
  year={2025},
  url={https://www.arxiv.org/abs/2511.00179}
}

@article{Ghanaatian_AI_submitted,
  title={Neural Network Based Molecular Structure Retrieval from Coulomb Explosion Imaging Data},
  author={Ghanaatian, Amir and Ravi, Aravinth K. and Stallbaumer, Joshua and Lam, Huynh V. S. and Rudenko, Artem and Greenman, Loren and Albin, Nathan and Caragea, Doina and Rolles, Daniel},
  journal={submitted},
  year={2025},
}

@article{Richard2025,
  author       = {B. Richard and R. Boll and S. Banerjee and J. M. Schäfer and Z. Jurek and G. Kastirke and K. Fehre and M. S. Schöffler and N. Anders and T. M. Baumann and S. Eckart and B. Erk and A. De Fanis and R. Dörner and S. Grundmann and P. Grychtol and M. Hofmann and M. Ilchen and M. Kircher and K. Kubicek and M. Kunitski and X. Li and T. Mazza and S. Meister and N. Melzer and J. Montano and V. Music and Y. Ovcharenko and C. Passow and A. Pier and N. Rennhack and J. Rist and D. E. Rivas and D. Rolles and I. Schlichting and L. Ph. H. Schmidt and P. Schmidt and D. Trabert and F. Trinter and R. Wagner and P. Walter and P. Ziolkowski and A. Rudenko and M. Meyer and R. Santra and L. Inhester and T. Jahnke},
  title        = {Imaging collective quantum fluctuations of the structure of a complex molecule},
  journal      = {Science},
  volume       = {389},
  number       = {6760},
  pages        = {650--654},
  year         = {2025},
  doi          = {10.1126/science.adk7912},
  url          = {https://www.science.org/doi/10.1126/science.adk7912}
}

@article{crane2023molecular,
  title={Molecular photodissociation dynamics revealed by {Coulomb} explosion imaging},
  author={Crane, Stuart W and Lee, Jason WL and Ashfold, Michael NR and Rolles, Daniel},
  journal={Physical Chemistry Chemical Physics},
  volume={25},
  number={25},
  pages={16672--16698},
  year={2023},
  publisher={Royal Society of Chemistry}
}

@article{burt2017coulomb,
  title={{Coulomb}-explosion imaging of concurrent {CH}$_2${B}r{I} photodissociation dynamics},
  author={Burt, Michael and Boll, Rebecca and Lee, Jason WL and Amini, Kasra and K{\"o}ckert, Hansjochen and Vallance, Claire and Gentleman, Alexander S and Mackenzie, Stuart R and Bari, Sadia and Bomme, C{\'e}dric and others},
  journal={Physical Review A},
  volume={96},
  number={4},
  pages={043415},
  year={2017},
  publisher={APS}
}

@article{Mazza_2023, title={The beam transport system for the Small Quantum Systems instrument at the European XFEL: optical layout and first commissioning results}, volume={30}, rights={https://creativecommons.org/licenses/by/4.0/}, ISSN={1600-5775}, DOI={10.1107/S1600577522012085}, abstractNote={The Small Quantum Systems instrument is one of the six operating instruments of the European XFEL, dedicated to the atomic, molecular and cluster physics communities. The instrument started its user operation at the end of 2018 after a commissioning phase. The design and characterization of the beam transport system are described here. The X-ray optical components of the beamline are detailed, and the beamline performances, transmission and focusing capabilities are reported. It is shown that the X-ray beam can be effectively focused as predicted by ray-tracing simulations. The impact of non-ideal X-ray source conditions on the focusing performances is discussed.}, number={2}, journal={Journal of Synchrotron Radiation}, author={Mazza, T. and Baumann, T. M. and Boll, R. and De Fanis, A. and Grychtol, P. and Ilchen, M. and Montaño, J. and Music, V. and Ovcharenko, Y. and Rennhack, N. and Rivas, D. E. and Rörig, A. and Schmidt, P. and Usenko, S. and Ziołkowski, P. and La Civita, D. and Vannoni, M. and Sinn, H. and Keitel, B. and Plönjes, E. and Jastrow, U. F. and Sorokin, A. and Tiedtke, K. and Mann, K. and Schäfer, B. and Breckwoldt, N. and Son, S.-K. and Meyer, M.}, year={2023}, month=mar, pages={457–467} }

@article{xgmd,
author = {Gruenert, Jan and Planas, Marc and Dietrich, Florian and Falk, Torben and Freund, Wolfgang and Koch, Andreas and Kujala, Naresh and Laksman, Joakim and Liu, Jia and Maltezopoulos, Theophilos and Tiedtke, Kai and Jastrow, Ulf and Sorokin, Andrey and Syresin, E. and Grebentsov, Alexander and Brovko, Oleg},
year = {2019},
month = {08},
pages = {1422-1431},
title = {X-ray photon diagnostics at the European XFEL},
volume = {26},
journal = {Journal of Synchrotron Radiation},
doi = {10.1107/S1600577519006611}
}
\bibliographystyle{rsc} %the RSC's .bst file

\end{document}

\pagestyle{fancy}
\thispagestyle{plain}
\fancypagestyle{plain}{
%%%HEADER%%%
\renewcommand{\headrulewidth}{0pt}
}
%%%END OF HEADER%%%

%%%PAGE SETUP - Please do not change any commands within this section%%%
\makeFNbottom
\makeatletter
\renewcommand\LARGE{\@setfontsize\LARGE{15pt}{17}}
\renewcommand\Large{\@setfontsize\Large{12pt}{14}}
\renewcommand\large{\@setfontsize\large{10pt}{12}}
\renewcommand\footnotesize{\@setfontsize\footnotesize{7pt}{10}}
\makeatother

\renewcommand{\thefootnote}{\fnsymbol{footnote}}
\renewcommand\footnoterule{\vspace*{1pt}% 
\color{cream}\hrule width 3.5in height 0.4pt \color{black}\vspace*{5pt}} 
\setcounter{secnumdepth}{5}

\makeatletter 
\renewcommand\@biblabel[1]{#1}            
\renewcommand\@makefntext[1]% 
{\noindent\makebox[0pt][r]{\@thefnmark\,}#1}
\makeatother 
\renewcommand{\figurename}{\small{Fig.}~}
\sectionfont{\sffamily\Large}
\subsectionfont{\normalsize}
\subsubsectionfont{\bf}
\setstretch{1.125} %In particular, please do not alter this line.
\setlength{\skip\footins}{0.8cm}
\setlength{\footnotesep}{0.25cm}
\setlength{\jot}{10pt}
\titlespacing*{\section}{0pt}{4pt}{4pt}
\titlespacing*{\subsection}{0pt}{15pt}{1pt}
%%%END OF PAGE SETUP%%%

%%%FOOTER%%%
\fancyfoot{}
\fancyfoot[LO,RE]{\vspace{-7.1pt}\includegraphics[height=9pt]{head_foot/LF}}
\fancyfoot[CO]{\vspace{-7.1pt}\hspace{11.9cm}\includegraphics{head_foot/RF}}
\fancyfoot[CE]{\vspace{-7.2pt}\hspace{-13.2cm}\includegraphics{head_foot/RF}}
\fancyfoot[RO]{\footnotesize{\sffamily{1--\pageref{LastPage} ~\textbar  \hspace{2pt}\thepage}}}
\fancyfoot[LE]{\footnotesize{\sffamily{\thepage~\textbar\hspace{4.65cm} 1--\pageref{LastPage}}}}
\fancyhead{}
\renewcommand{\headrulewidth}{0pt} 
\renewcommand{\footrulewidth}{0pt}
\setlength{\arrayrulewidth}{1pt}
\setlength{\columnsep}{6.5mm}
\setlength\bibsep{1pt}
%%%END OF FOOTER%%%

%%%FIGURE SETUP - please do not change any commands within this section%%%
\makeatletter 
\newlength{\figrulesep} 
\setlength{\figrulesep}{0.5\textfloatsep} 

\newcommand{\topfigrule}{\vspace*{-1pt}% 
\noindent{\color{cream}\rule[-\figrulesep]{\columnwidth}{1.5pt}} }

\newcommand{\botfigrule}{\vspace*{-2pt}% 
\noindent{\color{cream}\rule[\figrulesep]{\columnwidth}{1.5pt}} }

\newcommand{\dblfigrule}{\vspace*{-1pt}% 
\noindent{\color{cream}\rule[-\figrulesep]{\textwidth}{1.5pt}} }

\makeatother
%%%END OF FIGURE SETUP%%%

%%%TITLE, AUTHORS AND ABSTRACT%%%
\twocolumn[

%%% xxx 
%   \begin{@twocolumnfalse}
% {\includegraphics[height=30pt]{head_foot/PCCP}\hfill\raisebox{0pt}[0pt][0pt]{\includegraphics[height=55pt]{head_foot/RSC_LOGO_CMYK}}\\[1ex]
% \includegraphics[width=18.5cm]{head_foot/header_bar}}\par
% \vspace{1em}
% \sffamily
%%% xxx -commented out by KB 2025-01-31

\begin{tabular}{m{4.5cm} p{13.5cm} }

\includegraphics{head_foot/DOI} & \noindent\LARGE{\textbf{
Supplementary Material for Probing the structure of cyclic hydrocarbon molecules with X-ray-induced Coulomb explosion imaging
$^\dag$}} \\%Article title goes here ...
\vspace{0.3cm} & \vspace{0.3cm} \\

 & \noindent\large{
Kurtis D.~Borne\textit{$^{a,b}$}, 
Rebecca Boll\textit{$^{c}$}, 
Thomas M.~Baumann\textit{$^{c}$}, 
Surjendu Bhattacharyya\textit{$^{a,b}$},
Martin Centurion\textit{$^{d}$},  
Keyu~Chen\textit{$^{a}$}, 
Benjamin Erk\textit{$^{e}$},
Alberto De Fanis\textit{$^{c}$},
Ruaridh Forbes\textit{$^{b,f}$},
Markus Ilchen\textit{$^{c,e,g}$},
Edwin Kukk\textit{$^{h}$},
Huynh V.~S.~Lam\textit{$^{a}$},
Xiang Li\textit{$^{b}$}, 
Lingyu Ma\textit{$^{i}$},
Tommaso~Mazza\textit{$^{c}$},
Michael Meyer\textit{$^{c}$},
Terence Mullins\textit{$^{c,e,l}$},
J.~Pedro F.~Nunes\textit{$^{d}$},
Asami Odate\textit{$^{i}$},
Shashank Pathak\textit{$^{a}$},
Daniel~Rivas\textit{$^{c}$},
Philipp Schmidt\textit{$^{c}$},
%Bjoern Senfftleben\textit{$^{c}$},
Florian~Trinter\textit{$^{j}$},
Sergey~Usenko\textit{$^{c}$},
Anbu S.~Venkatachalam\textit{$^{a}$},
Enliang Wang\textit{$^{a,k}$},
Peter M.~Weber\textit{$^{i}$}, 
Till~Jahnke\textit{$^{c}$},
Artem~Rudenko\textit{$^{a}$}, 
and Daniel~Rolles\textit{$^{a,*}$}
}
 \\%Author names go here ...

\end{tabular}

% \end{@twocolumnfalse} 
 \vspace{0.6cm}
]
%%%END OF TITLE, AUTHORS AND ABSTRACT%%%

%%%FONT SETUP - please do not change any commands within this section
\renewcommand*\rmdefault{bch}\normalfont\upshape
\rmfamily
\section*{}
\vspace{-1cm}

%%%FOOTNOTES%%%

\footnotetext{\textit{$^{a}$~J.\ R.~Macdonald Laboratory, Department of Physics, Kansas State University, Manhattan, KS, USA}}
\footnotetext{\textit{$^{b}$~Linac Coherent Light Source, SLAC National Accelerator Laboratory, Menlo Park, CA, USA}}
\footnotetext{\textit{$^{c}$~European XFEL, Schenefeld, Germany}} %22869 Schenefeld
\footnotetext{\textit{$^{d}$~Department of Physics and Astronomy, University of Nebraska–Lincoln, Lincoln, NE, USA}}
\footnotetext{\textit{$^{e}$~Deutsches Elektronen-Synchrotron DESY, Hamburg, Germany}}
\footnotetext{\textit{$^{f}$~Department of Chemistry, University of California, Davis, CA, USA}}
\footnotetext{\textit{$^{g}$~Department of Physics, University of Hamburg, Hamburg, Germany}}
\footnotetext{\textit{$^{h}$~Department of Physics and Astronomy, University of Turku, Turku, Finland}}
\footnotetext{\textit{$^{i}$~Department of Chemistry, Brown University, Providence, RI, USA}} %Rhode Island
\footnotetext{\textit{$^{j}$~Molecular Physics, Fritz-Haber-Institut der Max-Planck-Gesellschaft, Berlin, Germany}}
\footnotetext{\textit{$^{k}$~Hefei National Research Center for Physical Sciences at the Microscale and Department of Modern Physics, University of Science and Technology of China, Hefei, China}} %Hefei 230026,
\footnotetext{\textit{$^{l}$~The Hamburg Centre for Ultrafast Imaging, Universität Hamburg, Hamburg, Germany}}

%%%MAIN TEXT%%%%

%\section{Raw Data}

     \begin{figure*}
        \centering
        
        \includegraphics[width=2.2\columnwidth]{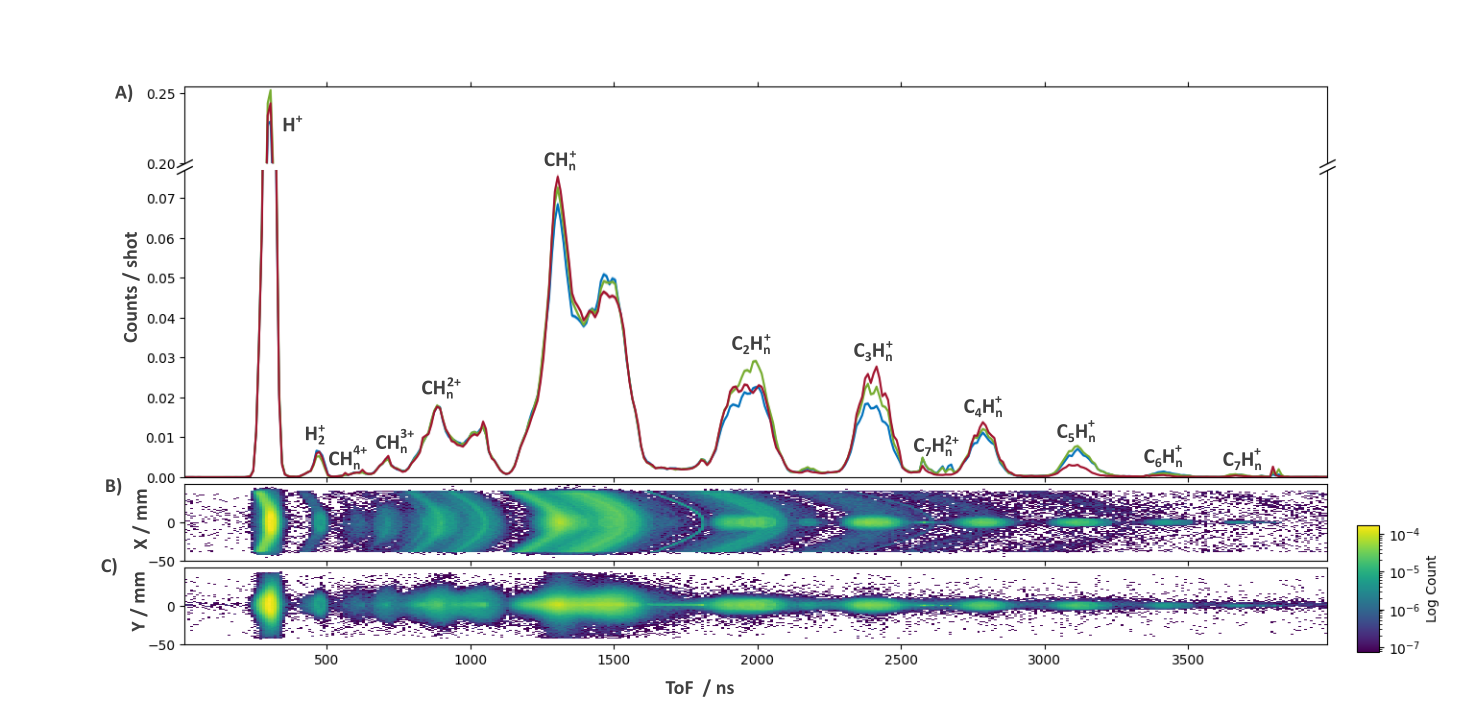}
        \caption{
        A) Ion time-of-flight spectrum for the three different isomers after ionization by 1.5-keV XFEL pulses with an average single-shot pulse energy of 4 mJ. 
        The subscript $n$ denotes the fact that each peak consists of several overlapping contributions from fragments containing a different number of hydrogen atoms, including $n=0$.
        For the \ce{CH_n^2+} and \ce{CH_n^+} peaks, the strongest contribution on the left-hand-side of each peak is from \ce{C^2+} and \ce{C^+}, respectively. 
        B) and C) Ion yield as a function of time-of-flight and hit position for toluene.}
        
        \label{fgr:tof_spec_XYT}
    \end{figure*}

   % \begin{figure*}
   %      \centering
        
   %      \includegraphics[height=10cm]{figures/pos_tof.png}
   %      \caption{Ion yield as a function of time-of-flight and hit position for toluene.}
        
   %      \label{fgr:pos_tof}
   %  \end{figure*}

   % \begin{figure*}
   %      \centering
        
   %      \includegraphics[width=2\columnwidth]{figures/detector_images.png}
   %      \caption{TOF-vs.-position and ion detector images of \ce{C^2+} fragments for toluene (panels A-C), cycloheptatriene (D-F), and 1,6-heptadiyne (G-I) for the triple-coincidence channel which three detected \ce{C^2+} ions.}
        
   %      \label{fgr:detector_images}
   %  \end{figure*}

   %     \begin{figure*}
   %      \centering
        
   %      \includegraphics[width=2\columnwidth]{figures/momentum_spheres.png}
   %      \caption{\ce{C^2+} ion momentum images for toluene (panels A-C), cycloheptatriene (D-F), and 1,6-heptadiyne (G-I) for the triple-coincidence channel which three detected \ce{C^2+} ions.}
        
   %      \label{fgr:momentum_spheres}
   %  \end{figure*}

       \begin{figure*}
        \centering
        
        \includegraphics[width=\columnwidth]{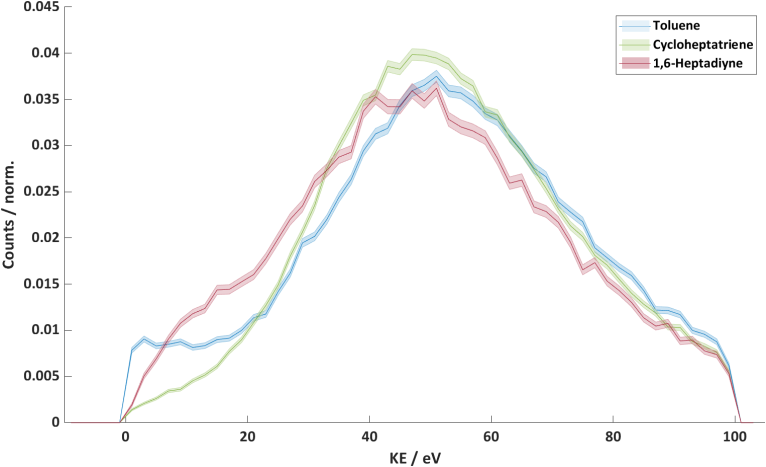}
        \caption{Experimental kinetic-energy spectrum of the C$^{2+}$ ions when selecting a coincidence of three C$^{2+}$ ions.}
        
        \label{fgr:tol_det_imag_C2+}
    \end{figure*}

   \begin{figure*}
        \centering
        
        \includegraphics[width=2\columnwidth]{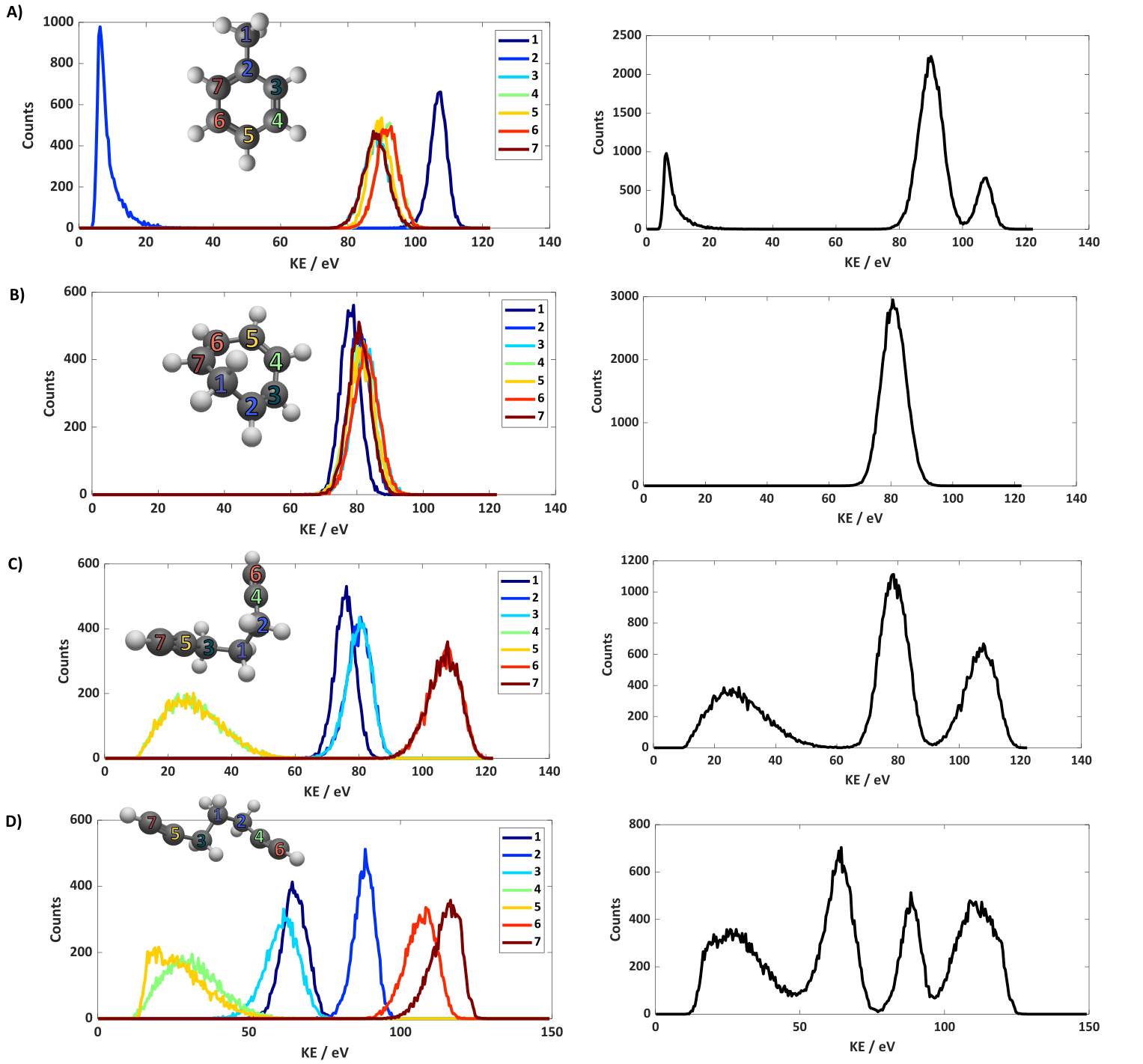}
        \caption{Simulated kinetic-energy spectra for a charge of 2+ on all carbon fragments for
        A) toluene, B) cycloheptatriene, C) the GG-trans conformer of 1,6-heptadiyne, and D) the AG conformer of 1,6-heptadiyne. The individual kinetic-energy spectra for each carbon site are color-coded in the left column, while the panels on the right show the total spectrum with the contributions from all carbons added. The corresponding experimental spectra are shown in Supplementary Fig.~2.}
        
        \label{fgr:KE_spectrum_simulated}
    \end{figure*}

   \begin{figure*}
        \centering
        
        \includegraphics[width=2\columnwidth]{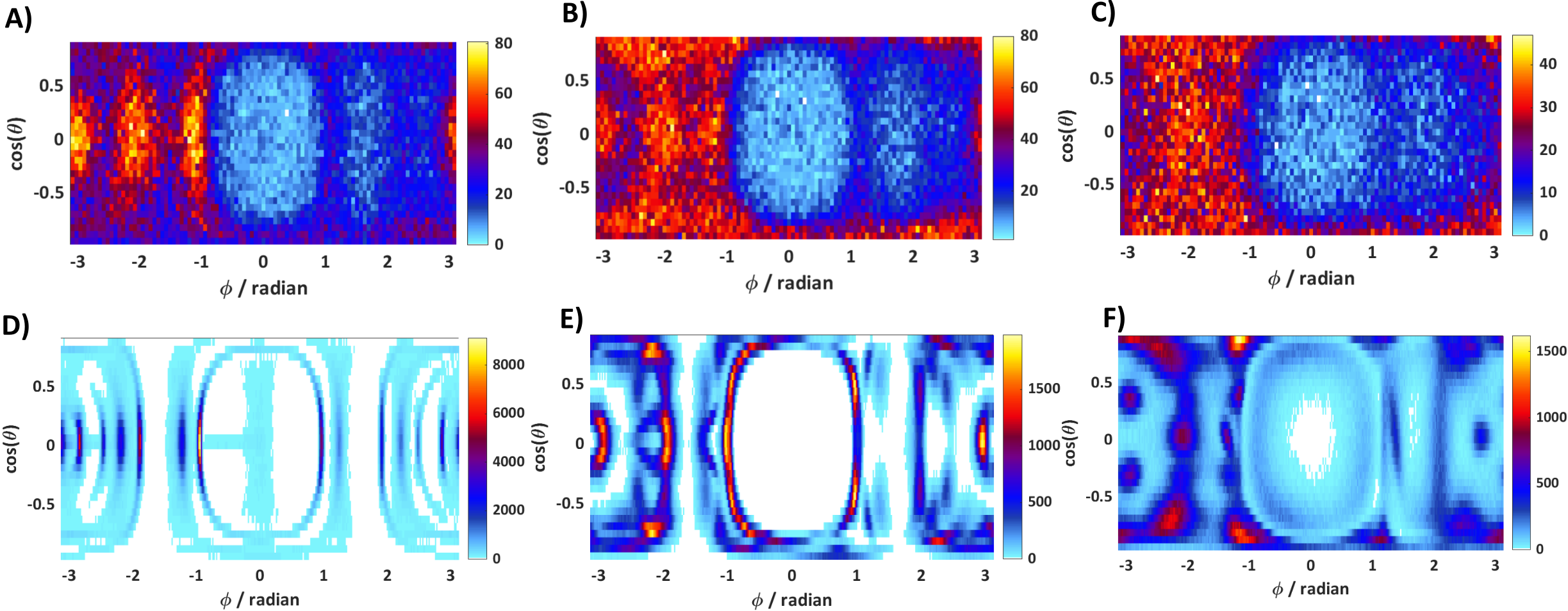}
        \caption{Experimental (top row) and simulated (bottom row) molecular-frame angular emission distributions in spherical coordinates for
        for C$^{2+}$ ions when combining all reference ions.
        A) and D) Toluene;
        B) and E) cycloheptatriene;
        C) and F) 1,6-heptadiyne (simulation contains only the GG-trans conformer geometry).
        }
        
        \label{fgr:Newton_plot_angles}
    \end{figure*}

   \begin{figure*}
        \centering
        
        \includegraphics[width=2\columnwidth]{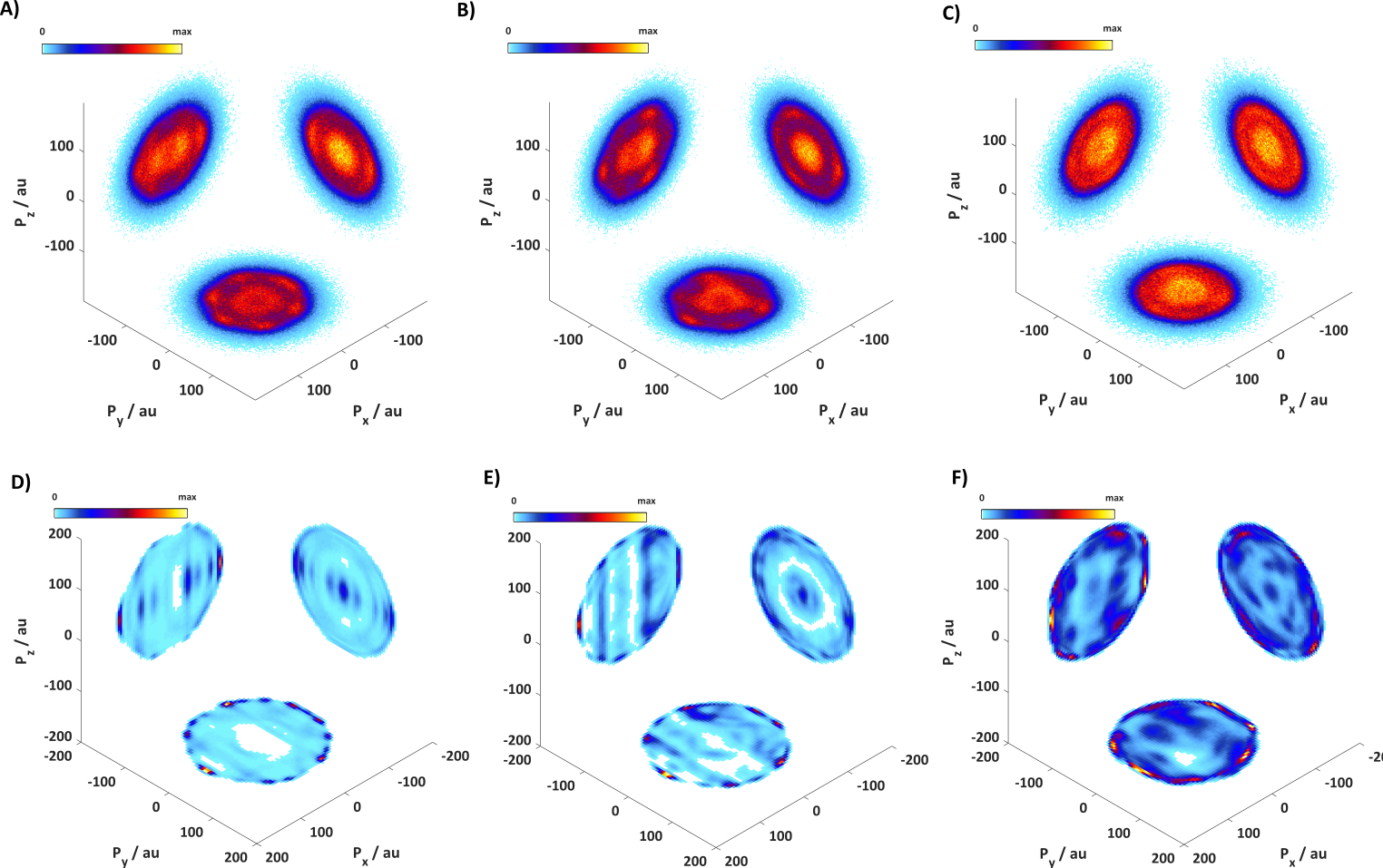}
        \caption{Experimental (top row) and simulated (bottom row) Newton plots for H$^+$ ions from the H$^+$ + C$^{2+}$ + C$^{2+}$ coincidence channel, with the recoil frame defined by the two C$^{2+}$ ions, when combining all reference ions.
        A) and D) Toluene;
        B) and E) cycloheptatriene;
        C) and F) 1,6-heptadiyne (simulation contains only GG-trans conformer geometry). 
        }
        
        \label{fgr:Newton_plot_angles}
    \end{figure*}

  \begin{figure*}
        \centering
        
        \includegraphics[width=2\columnwidth]{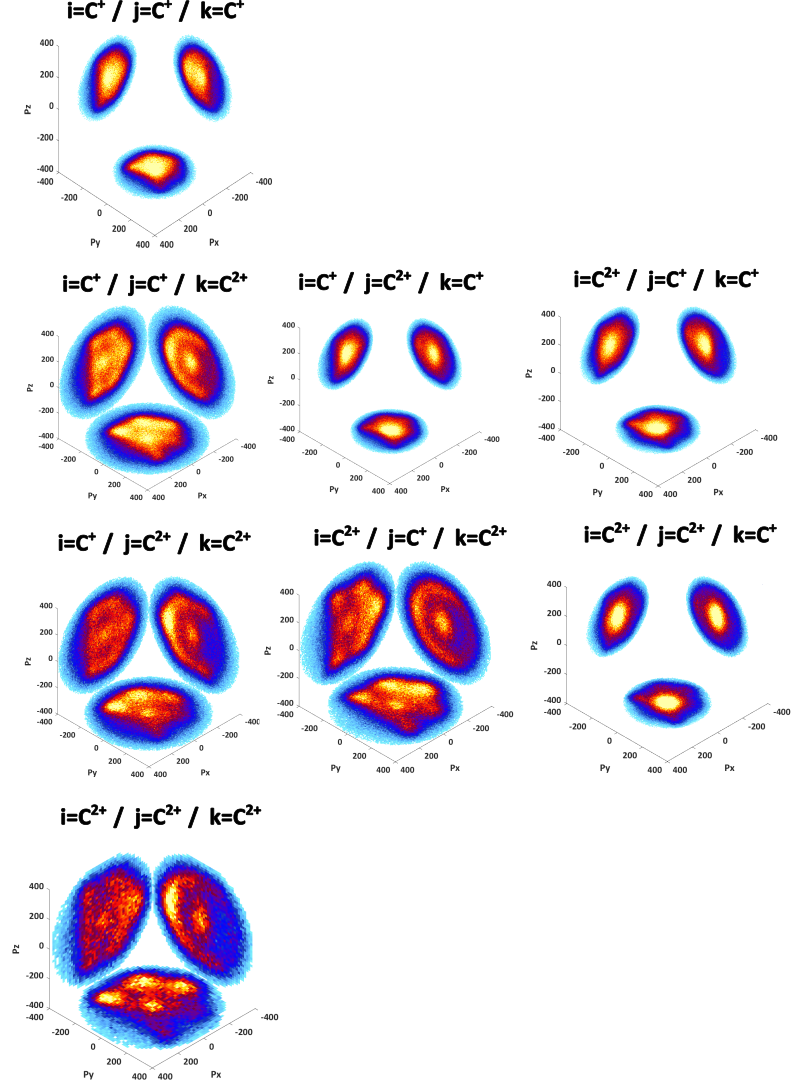}
        \caption{Experimental Newton plots for toluene from different charge channels and iterations of the recoil frame from the detected ions.}
        
        \label{fgr:toluene_charge_states}
    \end{figure*}

       \begin{figure*}
        \centering
        
        \includegraphics[width=2\columnwidth]{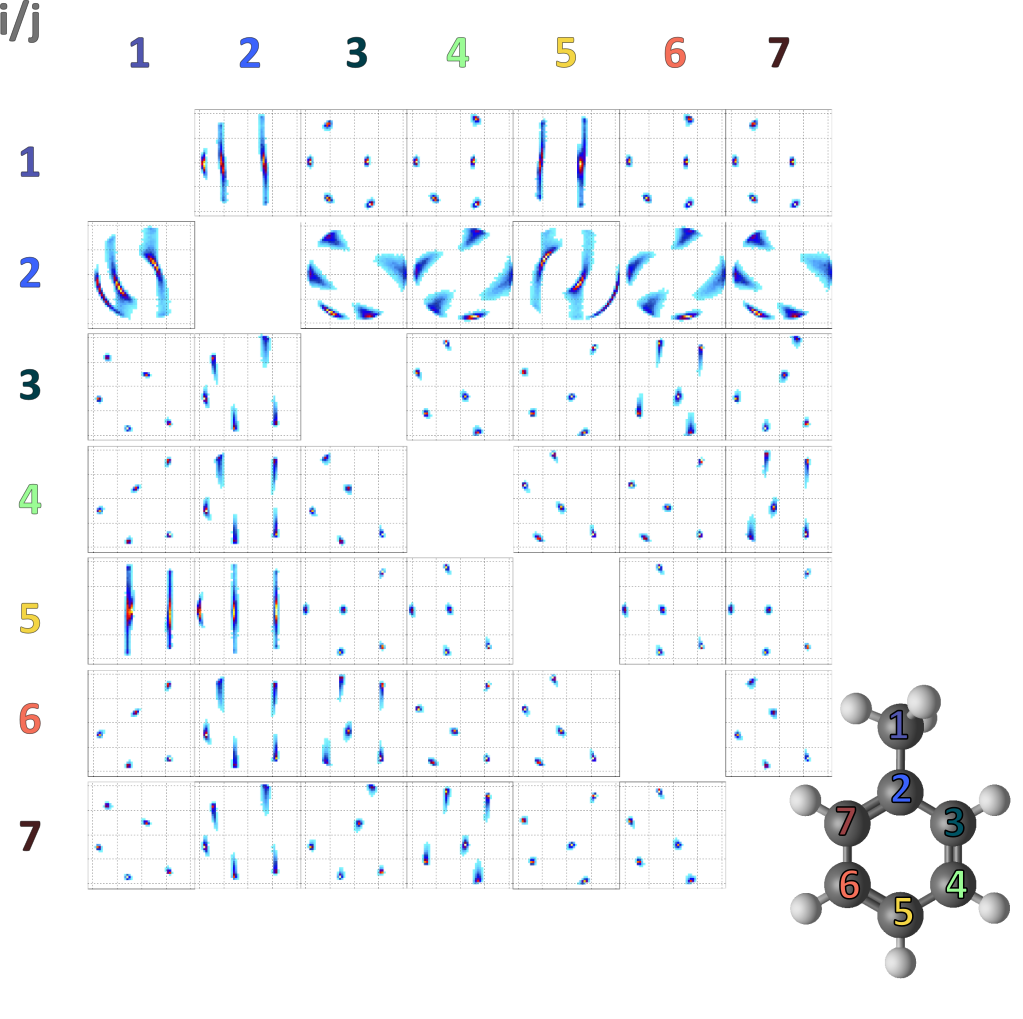}
        \caption{Simulated Newton plots for toluene iterating through all combinations of reference ions. Each column corresponds to the carbon site whose momenta defines the x-axis. The rows correspond to the carbon site whose momentum vector define the y-axis.}
        
        \label{fgr:toluene_reference_matrix}
    \end{figure*}

       \begin{figure*}
        \centering
        
        \includegraphics[width=2\columnwidth]{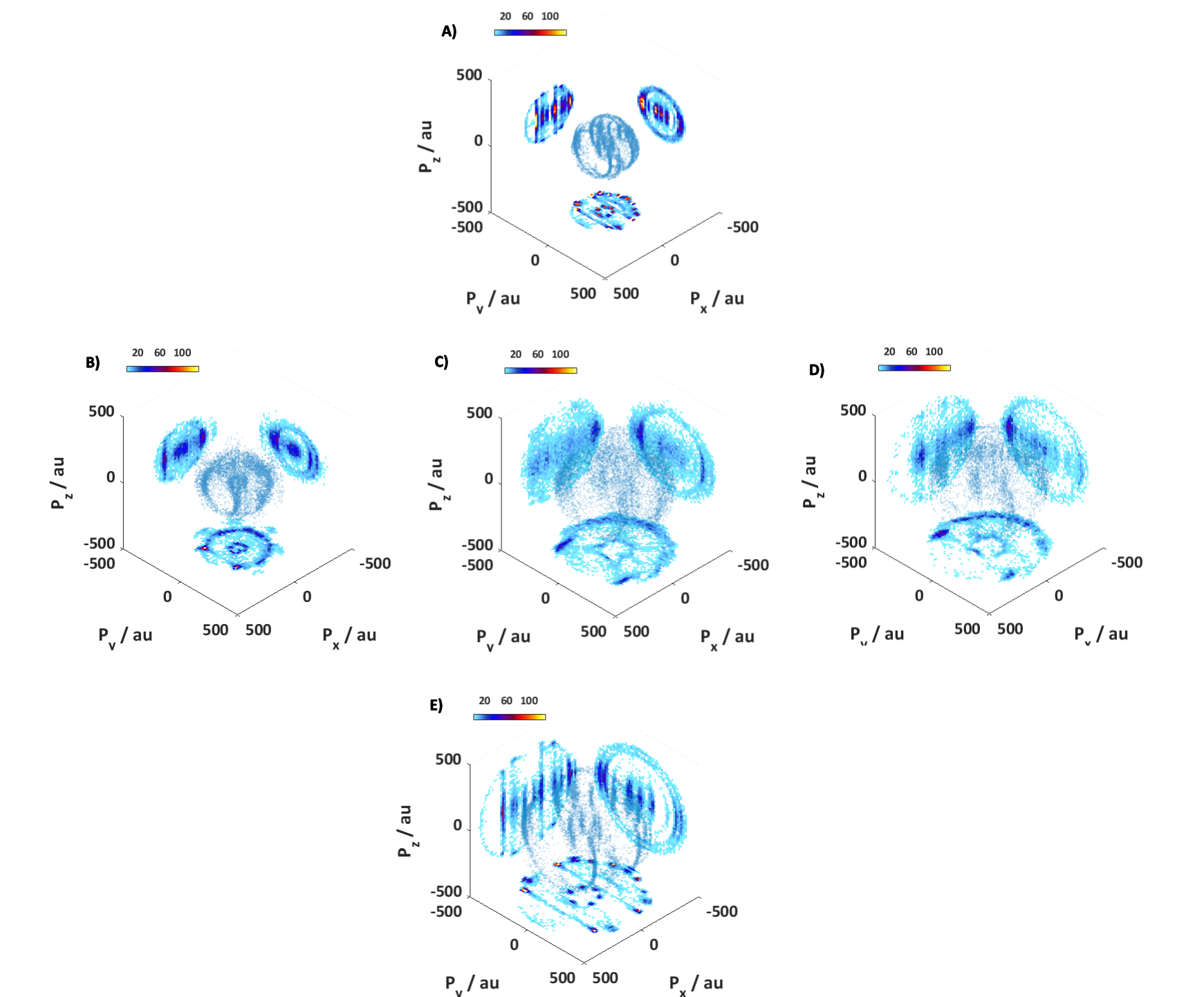}
        \caption{Simulated Newton plots for toluene from different charge channels and iterations of the recoil frame from the detected ions.
        A) Example of uniform charge distribution with each carbon having $q=+1$.
        B) Random charge distribution, where three fragments, each with a charge of $q=+1$, were selected from the ensemble. 
        C) Random charge distribution but with asymmetric selection of the fragments. Two fragments with $q=+1$ were selected, and a third with $q=+2$; the C$^{+}$ fragments define the recoil plane, while the C$^{2+}$ ion is plotted.
        D) Random charge distribution, where three fragments, each with a charge of $q=+2$, were selected from the ensemble.
        E) Uniform charge distribution with each carbon ion having a charge of $q=+2$.}
        
        \label{fgr:toluene_charge_states_sim+}
    \end{figure*}

%%%END OF MAIN TEXT%%%


\balance


%%%REFERENCES%%%
%\bibliography{rsc} %You need to replace "rsc" on this line with the name of your .bib file
%\bibliographystyle{rsc} %the RSC's .bst file